%% file: main.tex
\documentclass[journal]{IEEEtran}

\ifCLASSOPTIONcompsoc
  \usepackage[caption=false,font=normalsize,labelfont=sf,textfont=sf]{subfig}
\else
  \usepackage[caption=false,font=footnotesize]{subfig}
\fi

\input{preamble}

\begin{document}
\frenchspacing

\title{Influence of Sensor Feedback Limitations on \\ Power Oscillation Damping and Transient Stability
}
\author{
	Joakim~Bj\"ork${\textsuperscript{\orcidicon{0000-0003-0656-7991}}}$,~\IEEEmembership{Member,~IEEE,}
	Danilo Obradovi\'{c}${\textsuperscript{\orcidicon{0000-0001-6541-7892}}}$,~\IEEEmembership{Student~Member,~IEEE,}%
	\\
	Lennart~Harnefors${\textsuperscript{\orcidicon{0000-0002-3107-7073}}}$,~\IEEEmembership{Fellow,~IEEE,}
	and~Karl~Henrik~Johansson${\textsuperscript{\orcidicon{0000-0001-9940-5929}}}$,~\IEEEmembership{Fellow,~IEEE}
\thanks{This work was supported by the KTH PhD program in the digitalization of electric power engineering and in part by the Knut and Alice
	Wallenberg Foundation, the Swedish Research Council, the Swedish Foundation for Strategic Research, and by the multiDC project, funded by Innovation Fund Denmark, Grant Agreement No. 6154-00020B.
}%
	\thanks{J. Bj\"ork, D. Obradovi\'{c}, L. Harnefors, and K. H. Johansson are with the School of Electrical Engineering and Computer Science, KTH Royal Institute of Technology, \mbox{100 44 Stockholm,} Sweden (email: joakbj@kth.se; daniloo@kth.se; kallej@kth.se).}
	\thanks{L. Harnefors is also with ABB, Corporate Research, 721 78 V\"aster\aa s, Sweden  (email: lennart.harnefors@se.abb.com).}%
} 

\maketitle

\begin{abstract}
Fundamental sensor feedback limitations for improving rotor angle stability using local frequency or phase angle measurement are derived. 
Using a two-machine power system model, it is shown that improved damping of inter-area oscillations must come at the cost of reduced transient stability margins, regardless of the  control design method. 
The control limitations stem from that the excitation of an inter-area mode by external disturbances cannot be estimated with certainty using local frequency information.  
The results are validated on a modified Kundur four-machine two-area test system where the active power is modulated on an embedded high-voltage dc link. 
Damping control using local phase angle measurements, unavoidably leads to an increased rotor angle deviation following certain load disturbances. For a highly stressed system, it is shown that this may lead to transient instability. The limitations derived in the paper may motivate the need for wide-area measurements in power oscillation damping control.
\end{abstract}


\input{body}

\bibliographystyle{IEEEtran}
\bibliography{BibSensorFeedback}

\begin{IEEEbiography}[{\includegraphics[width=1in,height=1.25in,clip,keepaspectratio]{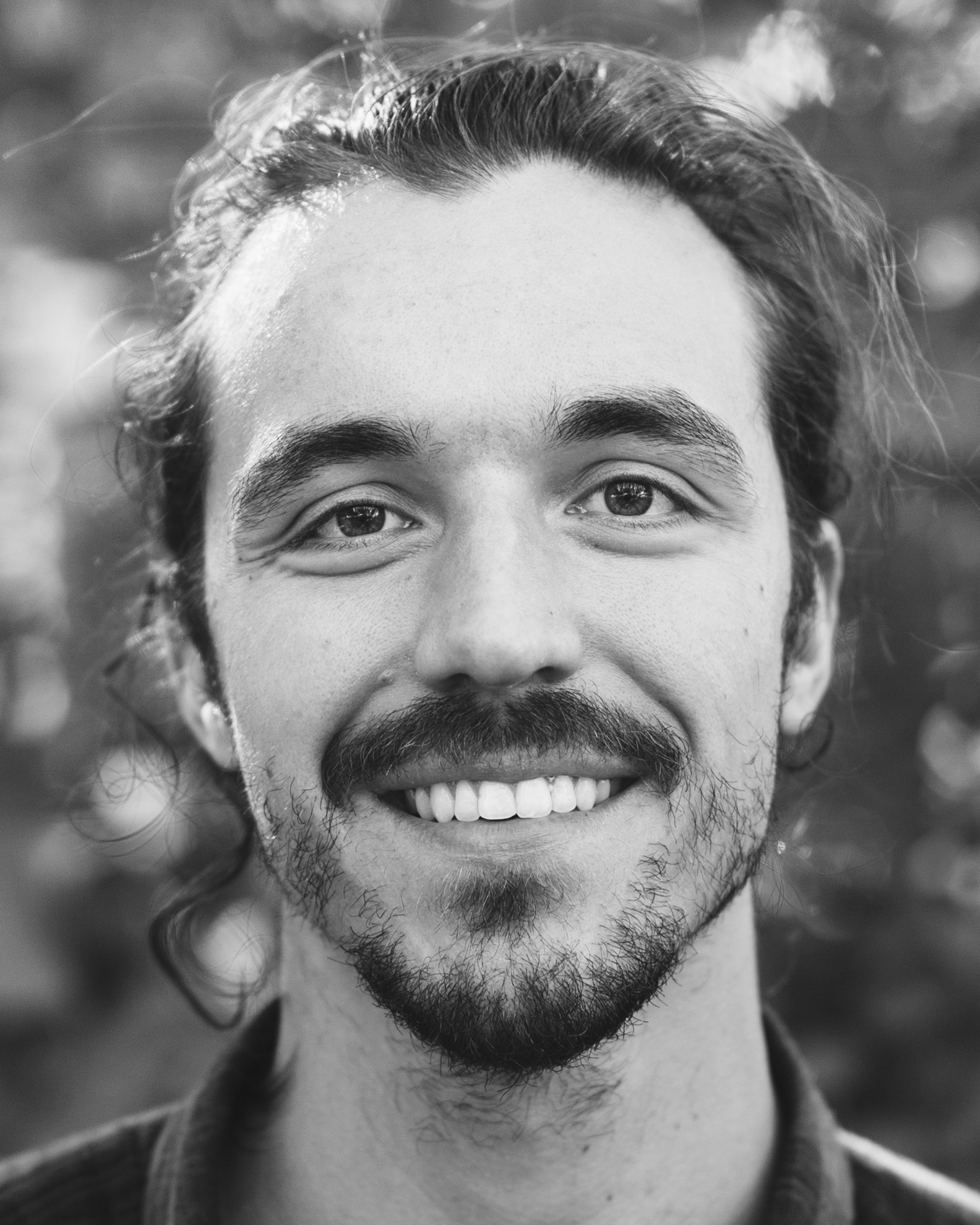}}]{Joakim Bj\"ork} 
	(S'17--M'21)
	received the M.Sc.\ degree in energy systems engineering from Uppsala University, Uppsala, Sweden, in collaboration with Swedish University of Agricultural Sciences, Uppsala, Sweden, in 2016.
	He obtained the Ph.D.\ degree in electrical engineering at the Division of Decision and Control Systems, KTH Royal Institute of Technology, Stockholm, Sweden, in 2021.
	
	In 2021, he joined the Swedish National Grid,
	Department of Power Systems. His research interests are fundamental control limitations in networked control systems with applications in frequency and voltage control of ac/dc power systems.
\end{IEEEbiography}

\begin{IEEEbiography}[{\includegraphics[width=1in,height=1.25in,clip,keepaspectratio]{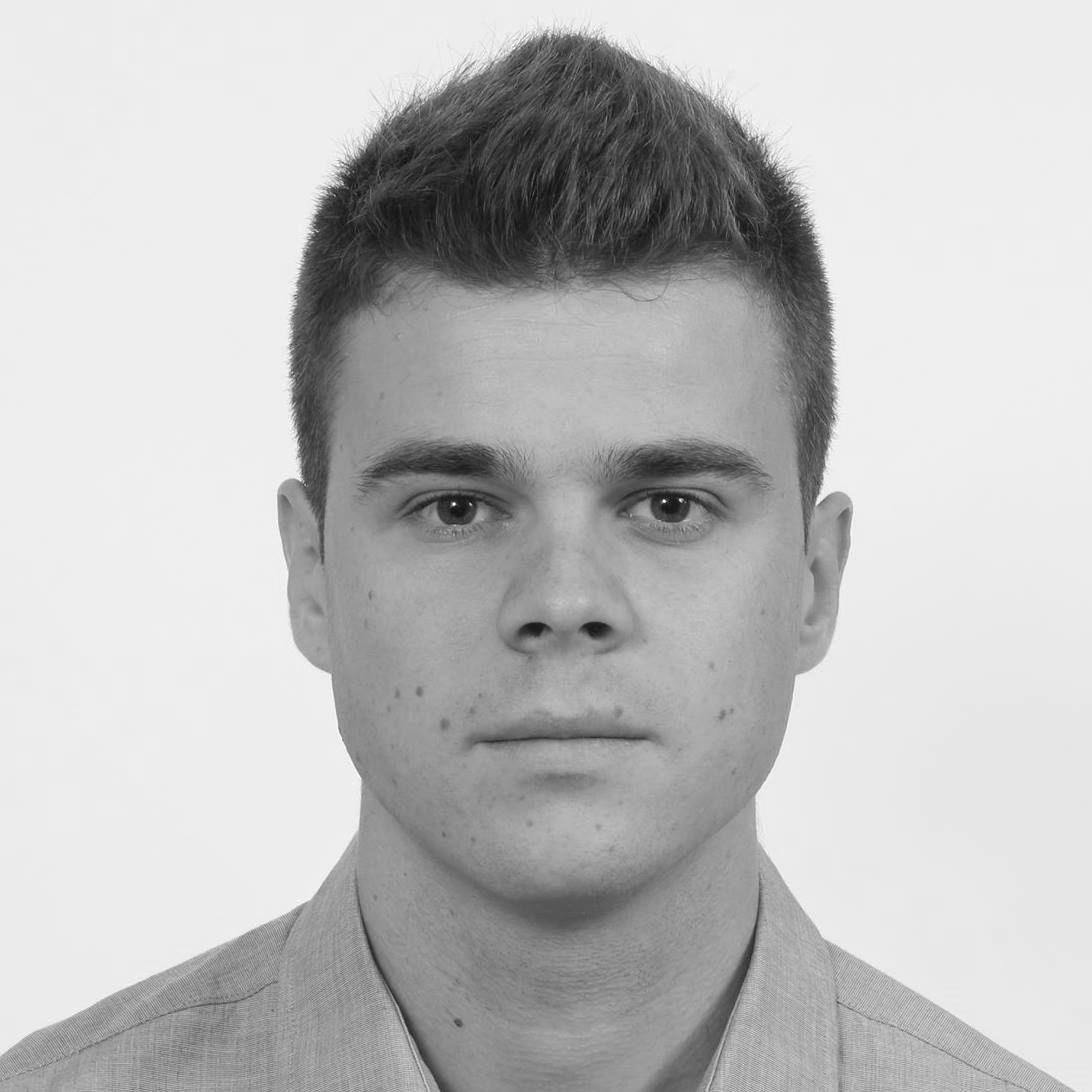}}]{Danilo Obradovi\'{c}} 
(S'18) received the B.Sc.\ and
M.Sc.\ degrees in electrical engineering at the Department of Power Systems from the School of Electrical
Engineering, University of Belgrade, Serbia, in 2016
and 2017, respectively. He obtained a Licentiate
degree in electrical engineering at the Division of
Electric Power and Energy Systems, KTH Royal Institute of Technology, Stockholm, Sweden, in 2020.

He is currently pursuing a Ph.D.\ degree with the
KTH Royal Institute of Technology. His research
interests include power system dynamics, stability
and control, and HVDC systems.
\end{IEEEbiography}

\vfill

\begin{IEEEbiography}[{\includegraphics[width=1in,height=1.25in,clip,keepaspectratio]{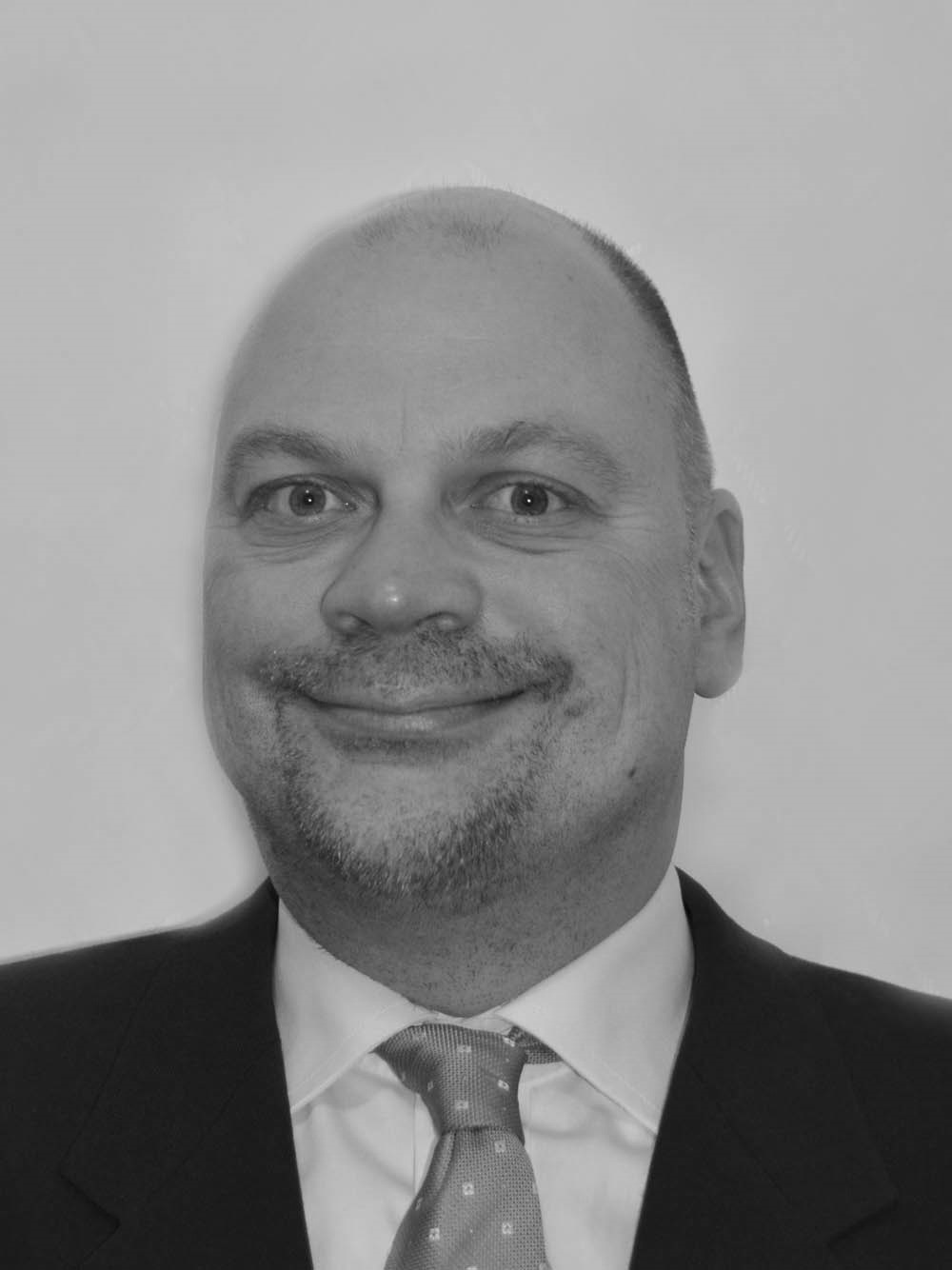}}]{Lennart Harnefors} 
	(F'17) 
	received the M.Sc.,	Licentiate, and Ph.D.\ degrees in electrical engineering from the Royal Institute of Technology (KTH), Stockholm,	Sweden, and the Docent (D.Sc.) degree in industrial	automation from Lund University, Lund, Sweden, in 1993,	1995, 1997, and 2000, respectively.
	
	From 1994 to 2005, he was with M\"alardalen University, V\"aster\aa s, Sweden, from 2001 as a Professor of electrical engineering. From 2001 to 2005, he was, in addition, a part-time Visiting Professor of electrical drives with Chalmers University of Technology, G\"oteborg, Sweden.
	
	In 2005, he joined ABB, HVDC Product Group, Ludvika, Sweden, where, among other duties, he led the control development of the first generation of	multilevel-converter HVDC Light. In 2012, he joined ABB, Corporate Research, V\"aster\aa s, where he was appointed as a Senior Principal Scientist in 2013 and as a Corporate Research Fellow in 2021. He is,	in addition, a part-time Adjunct Professor of power electronics with KTH.
	
	Dr. Harnefors is an Editor of the IEEE Journal of Emerging and Selected Topics	in Power Electronics and an Associate Editor of IET Electric Power Applications.	He was the recipient of the 2020 IEEE Modeling and Control Technical Achieved Award and was acknowledged as an outstanding reviewer of IEEE Transactions of Power Electronics in 2018. His research interests include control and dynamic analysis of power electronic systems, particularly grid-connected converters and ac drives. 
\end{IEEEbiography}

\begin{IEEEbiography}[{\includegraphics[width=1in,height=1.25in,clip,keepaspectratio]{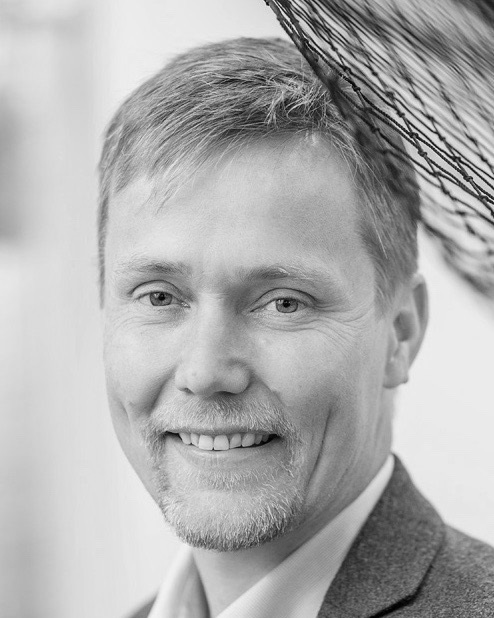}}]{Karl Henrik Johansson} 
	(F'13) 
	Karl H. Johansson is Professor with the School of Electrical Engineering and Computer Science at KTH Royal Institute of Technology in Sweden and Director of Digital Futures. He received M.Sc.\ and Ph.D.\ degrees from Lund University. 
	
	He has held visiting positions at UC Berkeley, Caltech, NTU, HKUST Institute of Advanced Studies, and NTNU. His research interests are in networked control systems and cyber-physical systems with applications in transportation, energy, and automation networks. He is a member of the Swedish Research Council's Scientific Council for Natural Sciences and Engineering Sciences. He has served on the IEEE Control Systems Society Board of Governors, the IFAC Executive Board, and is currently Vice-President of the European Control Association. He has received several best paper awards and other distinctions from IEEE, IFAC, and ACM. He has been awarded Distinguished Professor with the Swedish Research Council and Wallenberg Scholar with the Knut and Alice Wallenberg Foundation. He has received the Future Research Leader Award from the Swedish Foundation for Strategic Research and the triennial Young Author Prize from IFAC. He is Fellow of the IEEE and the Royal Swedish Academy of Engineering Sciences, and he is IEEE Control Systems Society Distinguished Lecturer.
\end{IEEEbiography}

\vfill

\end{document}

%% file: preamble.tex
\usepackage[pdftex]{graphicx}
\usepackage[usenames, dvipsnames]{color}
\usepackage[hidelinks]{hyperref} 

\usepackage[cmex10]{amsmath}
\usepackage{amssymb}    
\usepackage{bm}         
\usepackage{url}

\usepackage{mathtools} 
\usepackage{microtype}

\usepackage{steinmetz}
\usepackage{enumitem}

\usepackage{tikz}
\newcommand*\circled[1]{%
	\tikz[baseline=(C.base)]\node[draw,circle,inner sep=1.0pt,line width=0.2mm,](C) {$#1$};}
\newcommand*\Myitem{%
	\stepcounter{enumi}\item[\circled{\theenumi}]}

\usepackage{placeins}

\usepackage[nameinlink,capitalise]{cleveref}
\crefname{appsec}{Appendix}{Appendices}
\crefname{equation}{}{}
\usepackage{cite}

\usepackage{titlecaps}

\usepackage{accents}

\newcommand\munderbar[1]{%
	\underaccent{\bar}{#1}}

\newcommand{\HVDClink}{HVDC~link}

\newcommand{\Tconj}{\textit{\textsf{H}}}
\newcommand{\T}{\textit{\textsf{T}}}

\newcommand{\idx}{  \ensuremath{i} }
\newcommand{\jimag}{  \ensuremath{j} }
\newcommand{\jomega}{  \ensuremath{\jimag \omega} }

\newcommand{\inv}{\ensuremath{{-1}}}

\newcommand{\real}{\ensuremath{\mathbb{R}}}
\newcommand{\complex}{\ensuremath{\mathbb{C}}}

\DeclareMathOperator{\Real}{\ensuremath{\mathrm{Re}}}
\DeclareMathOperator{\Imag}{\ensuremath{\mathrm{Im}}}
\DeclareMathOperator{\diag}{\ensuremath{\mathrm{diag}}}
\DeclareMathOperator{\diagf}{\ensuremath{\mathrm{diag}}}

\newcommand{\filtP}{\ensuremath{\mathcal{P}}}
\newcommand{\filtM}{\ensuremath{\mathcal{M}}}
\newcommand{\sensS}{\ensuremath{\mathcal{S}}}
\newcommand{\sensT}{\ensuremath{\mathcal{T}}}
\newcommand{\Rzd}{\ensuremath{\mathcal{R}_{zd}}}
\newcommand{\Rzdone}{\ensuremath{\mathcal{R}_{zd_1}}}
\newcommand{\Rzdtwo}{\ensuremath{\mathcal{R}_{zd_2}}}
\newcommand{\Rzdk}{\ensuremath{\mathcal{R}_{zd_\idx}}}
\DeclareMathOperator{\sign}{\ensuremath{\mathrm{sgn}}}

\newcommand{\modelD}{\ensuremath{D_g}}
\newcommand{\modelM}{\ensuremath{M_g}}
\newcommand{\identity}{\ensuremath{1}}

\newcommand{\machinespeeds}{\ensuremath{\omega_g}}
\newcommand{\dotmachinespeeds}{\ensuremath{\dot{\omega}_g}}

\newtheorem{rem}{Remark}

\newtheorem{defi}{Definition}
\newtheorem{prop}{Proposition}

\newtheorem{lemma}{Lemma}

\newtheorem{ex}{Example}

\usepackage{siunitx}
\DeclareSIUnit\perunit{p.u.}
\DeclareSIUnit\voltampere{VA}

\usepackage{scalerel}
\usepackage{tikz}
\usetikzlibrary{svg.path}

\definecolor{orcidlogocol}{HTML}{A6CE39}
\tikzset{
	orcidlogo/.pic={
		\fill[orcidlogocol] svg{M256,128c0,70.7-57.3,128-128,128C57.3,256,0,198.7,0,128C0,57.3,57.3,0,128,0C198.7,0,256,57.3,256,128z};
		\fill[white] svg{M86.3,186.2H70.9V79.1h15.4v48.4V186.2z}
		svg{M108.9,79.1h41.6c39.6,0,57,28.3,57,53.6c0,27.5-21.5,53.6-56.8,53.6h-41.8V79.1z M124.3,172.4h24.5c34.9,0,42.9-26.5,42.9-39.7c0-21.5-13.7-39.7-43.7-39.7h-23.7V172.4z}
		svg{M88.7,56.8c0,5.5-4.5,10.1-10.1,10.1c-5.6,0-10.1-4.6-10.1-10.1c0-5.6,4.5-10.1,10.1-10.1C84.2,46.7,88.7,51.3,88.7,56.8z};
	}
}

\newcommand\orcidicon[1]{\href{https://orcid.org/#1}{\mbox{\scalerel*{
				\begin{tikzpicture}[yscale=-1,transform shape]
					\pic{orcidlogo};
				\end{tikzpicture}
			}{|}}}}

%% file: body.tex
\section{Introduction}
\IEEEPARstart{W}{ith} a steadily growing demand, deregulated market, and rising share of renewables the usage of long-distance power transfer is increasing. In heavily stressed grids, poorly damped inter-area modes are a concern \cite{kundurDefinitionClassificationPower2004}. A historical example is the western North American blackout of August 10, 1996, resulting from the instability of a \SI{0.25}{\hertz} inter-area mode \cite{venkatasubramanianAnalysis1996Western2004}. 
To improve the stability margins, and thereby increasing the transmission capacity,  power oscillation damping (POD) control is often implemented. Typical actuators are the excitation systems of generators via power system stabilizers (PSSs) \cite{kundurPowerSystemStability1994,kundurApplicationPowerSystem1989}, 
high-voltage direct current (HVDC) links \cite{pierreDesignPacificDC2019,trudnowskiPDCIDampingControl2013,smedUtilizingHVDCDamp1993,harneforsImpactInterareaModes2017,bjorkFundamentalControlPerformance2021,bjorkFundamentalPerformanceLimitations2019}, flexible alternating current transmission system (FACTS) devices \cite{martinsEfficientMethodsFinding1992,larsenSVCControlDesign1987,larsenConceptsDesignFACTS1995,kamwaWideareaMonitoringControl2006,zhangDesignWideareaPower2013,noroozianRobustControlStrategy2001,mhaskarPowerOscillationDamping2006}, inverter based power production \cite{dominguez-garciaControlSignalSelection2013}, and direct active power load control \cite{samuelssonDampingElectromechanicalOscillations1997}.

Traditionally, POD control is implemented as feedback control using locally available measurements.
Shortcomings of these measurements, such as potentially poor observability of system wide inter-area modes, can be circumvented by collecting measurements from distant geographical locations. With the increased number of installed phasor measurement units (PMUs), POD using wide-area measurement systems (WAMS) has become popular.

In \cite{kamwaWideareaMonitoringControl2006}, it is found that damping control based on specific local measurements may cause transient instability for some contingencies, whereas WAMS yield transient stability improvements as well as damping over a wide range of operating conditions. 
The implementation of a POD controller using WAMS and active power modulation of the Pacific HVDC Intertie in the western North American power system is presented in \cite{pierreDesignPacificDC2019}. Preliminary studies \cite{trudnowskiPDCIDampingControl2013} found local frequency measurement to be a suitable sensor signal for POD as it shows good observability and robust performance over a range of operating conditions. However, it is observed that damping based on local frequency measurements may deteriorate transient performance and cause first swing instability for some scenarios. The use of WAMS is found to improve performance also during severe contingencies \cite{trudnowskiPDCIDampingControl2013,pierreDesignPacificDC2019}.
Although providing unprecedented advantages for system monitoring and control, WAMS have some obvious drawbacks, such as, increased system complexity as well as potential reliability \cite{zhangDesignWideareaPower2013} and security issues \cite{adhikariWAMSCyberPhysicalTest2017}. Therefore, the use of WAMS should be well motivated.

Performance issues may be caused by the choice of measurements, but may also be a consequence of the control design. 
With optimization-based control design, good performance can often be achieved \cite{mondalPowerSystemSmall2014,abidoOptimalDesignPowersystem2002}. 
However, tuning of the optimization criteria can be an endless task as evaluating the achieved closed-loop performance is often far from trivial. 
Fundamental design limitations helps us to understand if unsatisfactory closed-loop performance---be it with traditional or modern control design methods---are due to a bad design or to inherent system limitations. 

The study of fundamental limitations in filtering and control design has a long history, dating back to the work of Bode in the 1940s \cite{bodeNetworkAnalysisFeedback1945}. Limitations associated with open right half-plane (RHP) poles and zeros as well as time-delays are quite well understood today.
For an overview of this area see~\cite{seronFundamentalLimitationsFiltering1997,freudenbergFrequencyDomainProperties1988,zhouRobustOptimalControl1996,skogestadMultivariableFeedbackControl2007,freudenbergFundamentalDesignLimitations2003,fangIntegratingControlInformation2017}.
These results has seen application in power systems. For instance in 
\cite{martinsEfficientMethodsFinding1992} where numerical methods to identify transfer function zeros in large power systems are presented and \cite{dominguez-garciaControlSignalSelection2013} where limitations due to RHP zeros are taken into account to identify a suitable input-output pairing to improve power system stability in a {general control configuration}. Typically, case studies of complex numerical models are performed to gain insight into the control problem at hand. However, resorting only to numerical solutions do not give valuable physical insight. Some notable exceptions are \cite{smedUtilizingHVDCDamp1993} where it is shown analytically on a two-machine power system how the location of controllable active power injections affects the potential of POD, \cite{samuelssonDampingElectromechanicalOscillations1997} where the positions of transfer function zeros are illustrated using an analogy with a damped mechanical pendulum, \cite{zhangInterconnectionTwoVery2011} where RHP zeros are identified in the connection of a voltage-source converter-based \HVDClink~to a weak ac system, and our  preliminary study \cite{bjorkControlLimitationsDue2020} where the destabilizing effect of automatic voltage regulators are identified as the cause for RHP zeros in a single-machine infinite bus system.

The choice of inputs and outputs can be motivated by studying pole-zero locations. The position of zeros, relative to the pole pair associated with the poorly damped inter-area mode, affects the controllability and observability of the mode. This holds true for RHP as well as for left half-plane (LHP) zeros. In power system small-signal analysis, controllability and observability are often combined into a single metric, the residue \cite{pagolaSensitivitiesResiduesParticipations1989}. The size of the residue reflects the effectiveness of the chosen input-output combination. The angle of the residue indicates the phase compensation required for a feedback controller to stabilize a poorly damped inter-area mode. Ideally, the residue and the control design should be insensitive to changing system conditions \cite{larsenSVCControlDesign1987} and fault locations \cite{noroozianRobustControlStrategy2001}. The position of zeros relative to  poorly damped poles are relevant since closed-loop poles tend to the position of nearby open-loop zeros with increasing feedback gain. Because of this, it is desirable to select a combination of input and output signals that results in a large separation between poles and zeros \cite{larsenConceptsDesignFACTS1995,mhaskarPowerOscillationDamping2006}. Although proven to be useful in practice, there is no straight-forward way to relate the notion of residue and pole-zero separation to the fundamental limitations of the closed-loop performance. In this work, we use the Bode integral constraint to bridge the gap between small-signal analysis and control performance limitations.
	
The contribution of this work is in the analysis of fundamental limitations for improving the overall rotor angle stability of the power system using local frequency measurements. 
Transient rotor angle stability is considered in terms of the system's ability to maintain synchronism in the first swing  following large load disturbances.
The considered system is characterized by the available control inputs and measurement signals, the desired performance variable, and the considered disturbances.
The control problem is formulated as a sensor feedback problem, partitioned into separate control and filtering problems.
The analysis is performed on a linearized two-machine power system representing a poorly damped inter-area mode. 
It is shown that with ideal measurements, performance in terms of rotor angle stability is only limited by the available input power.
Then it is shown, using Bode integral constraints on the filtering sensitivity, that the excitation of the inter-area mode cannot be accurately estimated using local voltage phase angle or frequency measurements. Last, it is shown that the consequence of this filtering limitation is that any damping improvement based on local frequency measurements, must come at the cost of reduced transient stability margins, as implied by the findings of \cite{kamwaWideareaMonitoringControl2006,trudnowskiPDCIDampingControl2013}. The results are validated in nonlinear simulations using the well-known Kundur four-machine two-area test system, where the active power of an embedded \HVDClink~is modulated to improve damping of the dominant inter-area mode. In the simulation study we extend the results by considering WAMS, local ac power flow measurement, and local bus voltage measurement.\footnote{The four-machine test system and the control design examples are available at the GitHub repository: \url{https://github.com/joakimbjork/4-Machine}.}

The remainder of this paper is organized as follows. In \cref{sec:Model} a linearized two-machine power system model is derived. \cref{sec:preliminaries} introduces results from the literature on sensor feedback limitations derived for general linear control and filtering problems. 
In \cref{sec:sensor_feedback_limitations} these results are applied to the two-machine power system model and explicit limitations are derived for two different sensor configurations.
In \cref{sec:Simulink} the results are validated on a nonlinear benchmark power system model. \cref{sec:conclusion} concludes the work.

\section{Linearized Power System Model}
\label{sec:Model}
In this section a linearized model of a two-machine power system is derived. A state-space representation is developed from physical equations based on simplifying assumptions. Then, transfer functions are derived, capturing the zero dynamics relevant for the control limitations studied in the paper.

The derived model assumes active power control only at the bus where we measure the voltage phase angles. In \cref{sec:Simulink} we assume modulation of active power in an \HVDClink~connecting two buses. Although the input controllability is improved compared to a single point of injection, this does not affect the sensor feedback limitations.

\subsection{State-Space Representation of a Two-Machine Power System}
\begin{figure}[t!]
	\centering
	\includegraphics[width=1\linewidth]{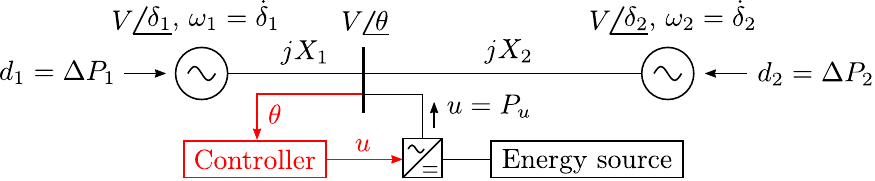}
	\caption{Two-machine power system  with a controlled supplementary active power injection. The energy source could for instance be a wind power park, a controllable load, or an \HVDClink~connecting to an asynchronous system.}
	\label{fig:system_2machine}
\end{figure}
We consider a power system represented by two synchronous machines connected by an ac transmission line shown in \cref{fig:system_2machine}.  Making simplifying assumptions in line with those of \cite{harneforsImpactInterareaModes2017} this can be considered a simplistic representation of the dominant inter-area mode in a more realistic multi-machine grid.

The electromechanical dynamics can be described using the swing equation 
\begin{equation}
\label{swing:eq1}
\begin{aligned}
\dot{\delta}_i &= \omega_i \\
M_i\dot{\omega}_i &= - \frac{V^2}{X_i}\sin(\delta_i-\theta) - D_i \omega_i + \Delta P_i 
\end{aligned}
\end{equation}
for $i = 1,2$.
Machines are modeled, using the classical machine model, as a stiff electromotive force behind a transient reactance~\cite{kundurPowerSystemStability1994}. 
Machine excitation and reactive power at the control bus are controlled so that all buses have constant voltages amplitudes $V$ for the time frame of interest. 
The rotor phase angles $\delta_i$
and machine speed  $\omega_i$ represents machine $i$'s deviation from a synchronously rotating reference frame with frequency {$2 \pi f_s$}, where $f_s$ is usually 50 or \SI{60}{Hz}.
Constants $M_i$ represents the frequency and pole-pair scaled inertia of each machine and $D_i$ represent the equivalent damping of higher-order dynamics such as impact from machine damper windings, voltage regulators, system loads, and governors etc.
The transmission is assumed lossless and the electrical distance between machine $i$ and the control bus is represented by the reactance $X_i$, consisting of transient machine reactance, transformers, and transmission lines.
The difference between the mechanical input power from the machines and the local loads is given by $\Delta P_i$.
Voltage phase angle $\theta$ at the control bus is given by the active-power balance
\begin{equation}
\label{AE1}
P_{u} + \sum_{i=1}^{2}\frac{V^2}{X_i}\sin(\delta_i - \theta)  = 0,
\end{equation}
where $P_{u}$ is active power injected at the control bus.

Linearizing around the operating point $\delta_i(t_0) = \delta_i^*$ and $\theta(t_0) =\theta^*$, we get approximately
\begin{equation}
\label{AE2}
\theta = \frac{X_2^*}{X_\Sigma^*}\delta_1 + \frac{X_1^*}{X_\Sigma^*}\delta_2 +\frac{X_1^*X_2^*}{X_\Sigma^*} P_u,
\end{equation}
where ${X_i^*}^\inv = X_i^\inv{V^2}\cos(\delta_i^*-\theta^*)$, and $X_\Sigma^* = X_1^*+X_2^*$. 
Substituting \eqref{AE2} into \eqref{swing:eq1} then gives us the linearized swing equation in state-space form
\begin{equation}
\label{swing:eq2}
\begin{aligned}
\begin{bmatrix}
\dot{  \delta}\\  \modelM { \dotmachinespeeds}
\end{bmatrix}  &= 
\begin{bmatrix}
0  & I \\
-L & -\modelD
\end{bmatrix} 
\begin{bmatrix}
\delta \\  \machinespeeds
\end{bmatrix}
+  \begin{bmatrix}
0 & 0 \\
I & L_\theta
\end{bmatrix}
\begin{bmatrix}
\Delta  P \\ P_u
\end{bmatrix},
\end{aligned}
\end{equation}
where input $\Delta  P = [\Delta P_1, \Delta P_2]^\T$, and state variables  $\delta = [\delta_1, \delta_2]^\T$ and $\machinespeeds  = [\omega_1, \omega_2]^\T$.\footnote{The notation $\omega$ (with no subscript) is reserved for angular frequency.}
Inertia and damping constants are given by $\modelM = \diagf(M_1,M_2)$ and $\modelD = \diagf(D_1,D_2)$, respectively, while $0$ and $I$ are appropriately sized zero and identity matrices, respectively. The linearized power flow are described by the network matrices 
\begin{equation}
\label{swing:eq2_L}
L = \frac{1}{X_\Sigma^*}\begin{bmatrix}
1 & -1\\ -1 & 1
\end{bmatrix}, \quad \text{and} \quad L_\theta = \frac{1}{X_\Sigma^*}\begin{bmatrix}
X_2^* \\ X_1^* 
\end{bmatrix}.
\end{equation}

\subsection{Transfer Function of a Two-Machine Power System}
\label{sec:Transfer_func_two-machine}
The sensor feedback limitations consider in this paper are connected with the controllability and observability of the inter-area mode. 
Commonly, the residue method \cite{kundurPowerSystemStability1994} is used to characterize the input-output controllability and observability of modes in small-signal analysis studies and POD controller design. 
However, for the purpose of this analysis, residues provide insufficient information. Instead we use \eqref{swing:eq2} to derive transfer functions capturing the poles and zeros of relevant input-output combinations.

The electrical midpoint may differ from the \textit{mass-weighted} electrical midpoint. However, to simplify notation we assume that the machines have identical inertia constants so that $M_1=M_2=M$. 
For convenience, we also assume that damping can be neglected so that $D_1=D_2=0$ and that we consider the extreme case where load disturbances $d = [d_1, d_2]^\T$ occur close to the machines so that $d_1 = \Delta P_1$ and $d_2 = \Delta P_2$. Consider $u = P_u$ in \cref{swing:eq2} to be a controlled active power injection somewhere between the two machines as shown in \cref{fig:system_2machine}.
The transfer function of \cref{swing:eq2} mapping external inputs $d$ and $u$ to phase angles at machine and control buses $\delta=
[\delta_1,\delta_2
]^\T$ and $\theta$, respectively, are given by
\begin{equation}
\label{eq:Gdeltaw}
\setlength\arraycolsep{2pt}\left[
\begin{array}{c|c}
G_{\delta d}& G_{\delta u}   \\ \hline G_{\theta d} & G_{\theta u} 
\end{array}\right]=G_0 \! \left[\begin{array}{cc|c}
\hspace{-1pt} {s^2+\Omega^2/2} \hspace{-1pt} & {\Omega^2/2} & N_1 \\
{\Omega^2/2} & \hspace{-1pt} {s^2+\Omega^2/2} \hspace{-1pt} & N_2 \\ \hline
N_1 & N_2 & N_3
\end{array}\right]\!,
\end{equation}
where
$\Omega = \sqrt{{2}/{MX_\Sigma^*}}$
is the undamped frequency of the inter-area mode and
\begin{equation}
\label{eq:G0_poles}
G_0  = \frac{1}{s^2 M\left(s^2+\Omega^2\right)}.
\end{equation}
In particular, for an unloaded system we have
\begin{equation}
\label{eq:Q_zeros}
N_1
= \frac{X_2^*}{X_\Sigma^*}\!
\left(\! s^2 + \frac{1}{MX_2^*}\!\right), \quad
N_2
= \frac{X_1^*}{X_\Sigma^*}\!
\left(\! s^2 +  \frac{1}{MX_1^*}\!\right),
\end{equation}
and $N_3 = MX_\Sigma^*N_1N_2$. 
\subsection{Mechanical Analogy}
\label{sec:mechanical_analogue}
\begin{figure}[tb!]
	\centering
	\includegraphics[width=\linewidth]{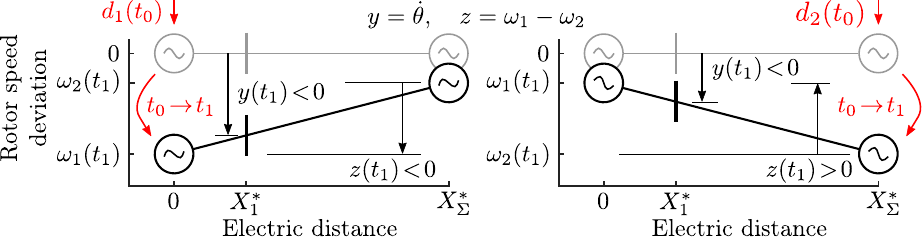}
	\caption{The system interpreted as two masses on a swing plank subject to an external force $d_1$ or $d_2$. At initial time $t_0$, $\omega_1(t_0) = \omega_2(t_0) = \dot \theta(t_0) = 0$. }
	\label{fig:system_2machine_problem}
\end{figure}

The machines in \cref{fig:system_2machine} can be interpreted as masses on a swing plank with the electrical
distance as physical distance. An external force applied to the system will initiate a relative swing between the two masses, if not applied directly at the center \cite{smedUtilizingHVDCDamp1993}. Now consider \cref{fig:system_2machine_problem}. We want to design a sensor feedback controller using local frequency (or phase angle) measurement, $y=\dot \theta$, to attenuate the relative swing. At time $t_0$, a force $d_1$ or $d_2$ is applied to one of the machines. To dampen the swing, the sign of $z = \omega_1-\omega_2$ need to be estimated. As seen in \cref{fig:system_2machine_problem}, however, this information is not directly attained from the local frequency measurement. A damping controller may therefore cause transient stability issues by amplifying the first swing as observed in \cite{trudnowskiPDCIDampingControl2013}. In this paper, we study the fundamental limitations of this control problem.

\section{Sensor Feedback Limitations}
\label{sec:preliminaries}
We review some results for general linear systems. In this paper, the scope is limited to scalar systems. For more extended overview see \cite{seronFundamentalLimitationsFiltering1997,freudenbergFrequencyDomainProperties1988,zhouRobustOptimalControl1996,skogestadMultivariableFeedbackControl2007,freudenbergFundamentalDesignLimitations2003,fangIntegratingControlInformation2017}. In \cref{sec:sensor_feedback_limitations} the result will be used to show sensor feedback limitations for the two-machine power system derived in \cref{sec:Model}.

Let the scalar transfer function $G_{yu}$ represent a plant with output $y$ and input $u$. Consider a linear controller
$
u = -Ky
$
where $K$ is any proper rational transfer function.

\begin{defi}
	[Sensitivity Functions]%
	The closed-loop sensitivity and complementary sensitivity functions are given by
	\begin{equation}
	\label{eq:SandT}
	\sensS = (\identity+G_{yu}K)^\inv, \quad \text{and} \quad \sensT = \identity-\sensS, 
	\end{equation}
	respectively.
\end{defi}

\begin{lemma}[Interpolation Constraints \cite{seronFundamentalLimitationsFiltering1997}]
	\label{lemma:Interpolation_constraints_on_S_and_T}%
	For internal stability, no cancellation of open RHP poles or zeros are allowed between the plant and the controller. Let $p_i,\, i= 1,\ldots, n_p$ and $q_i, \,  i=  1,\ldots, n_q$ be the open RHP poles and zeros of the plant $G_{yu}$. Then for all $p_i$ and $q_i$, 
	\begin{equation}
	\begin{aligned}
	\sensS(p_i) &= 0, \quad \sensT(p_i) = 1, \\ 
	\sensS(q_i) &= 1, \quad \sensT(q_i) = 0.
	\end{aligned} 
	\end{equation}
\end{lemma}

The interpolation constraints limit the achievable performance since $\sensS$ and $\sensT$ represent closed-loop amplification of load disturbances and measurement noise, respectively.

\begin{lemma}[Bode Integral \cite{seronFundamentalLimitationsFiltering1997}] 
	\label{lemma:Bode_sensitivity_integral}%
	Suppose that the loop-gain $G_{yu}K$ is a proper rational function.
	Then, if $\sensS(\infty)  \neq 0$ 
	\begin{multline}
	\label{eq:bodeInt_S}
	\int_{0}^{\infty }  \ln \left|\frac{\sensS( \jomega)}{\sensS(\infty)} \right| d\omega = \frac{\pi}{2}\lim\limits_{s\rightarrow \infty} \frac{s\left[\sensS(s)-\sensS(\infty)\right]}{\sensS(\infty)} 
	\\+ \pi \sum_{ \mathclap{i=1} }^{n_p} p_i\text{\normalsize }
	\end{multline}
	where $\jimag = \sqrt{-1}.$	
\end{lemma}

If the loop-gain is strictly proper, then $\sensS(\infty)  =1$. 
Typically, both $G_{yu}$ and $K$ are strictly proper.
The limit in  \eqref{eq:bodeInt_S} then goes to 0. 
The reduction of the sensitivity at some angular frequencies then has to be compensated by an (at least) equally large amplification in other  frequencies. This resembles the displacement of water in a water-bed. The Bode integral constraint is therefore also known as the water-bed effect.

\begin{figure}[t!]
	\centering
	\includegraphics[width=\linewidth]{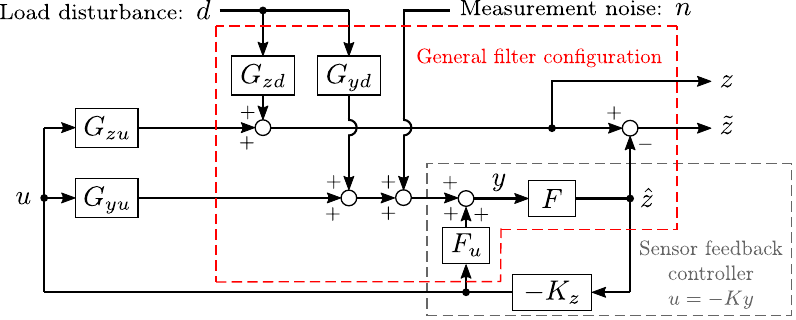}
	\caption{General control configuration as a control and filtering problem. The filter $F_u$ decouples the control input $u$ from the estimation error $\tilde z$. 
	}
	\label{fig:sensor_feedback}
\end{figure}

In general the measured output $y$ may differ from the performance variable $z$ that we want to control. 
For an open-loop plant in the general control configuration shown in 
\cref{fig:sensor_feedback},
the goal is to design a sensor feedback controller $K$ that reduces the amplitude of the closed-loop system from $d$ to $z$, given by
\begin{equation}
\label{eq:Tzd}
\sensT_{zd} = G_{zd} - G_{zu}K(\identity + G_{yu}K)^\inv  G_{yd}.
\end{equation}
In the general control configuration, RHP poles and zeros in $G_{yd},$ $G_{zd}$, and $G_{zu}$ may put further restrictions on the achievable performance \cite{freudenbergFundamentalDesignLimitations2003}.
The general control problem can be separated into a control and an estimation problem as shown in \cref{fig:sensor_feedback}. This work focuses on the latter by studying limitations in the linear filtering problem. 

Assume that the system is detectable from $y$, i.e., all unobservable states are stable, and that $\hat z = Fy$ is an unbiased, bounded error estimate of $z$.
An observer is a bounded error estimator if for all finite initial states, the estimation error $\tilde z = z-\hat z$ is bounded for all bounded inputs.
A bounded error estimator is unbiased if $u$ is decoupled from $\tilde z$ \cite{seronFundamentalLimitationsFiltering1997}. 

\begin{defi}[Filtering Sensitivity Functions \cite{seronFundamentalLimitationsFiltering1997}]
	\label{defi:Filtering_Sensitivity}%
	If $G_{yd}$ is detectable, $F$ is a stable filter, and $G_{zd}$ is right invertible\footnote{For $G_{zd}$ to be right invertible there need to be at least as many inputs as signals to be estimated. Note that $G_{zd}^\inv$ is not necessarily proper.},
	the filtering sensitivity functions are given by
	\begin{equation}
	\label{eq:P_and_M}
	\filtP = (G_{zd} -FG_{yd})G_{zd}^\inv, \quad \text{and} \quad \filtM = FG_{yd}G_{zd}^\inv,
	\end{equation}
	with $
	\filtP(s) + \filtM(s) = \identity
	$
	at any $s\in \complex$ that is not a pole of $\filtP$ or $\filtM$.	
\end{defi}

The filtering sensitivity function $\filtP$ represents the relative effect of disturbance $d$ on the estimation error $\tilde z$, while the complementary filter sensitivity $\filtM$ represents the relative effect of $d$ on the estimate $\hat z$.

\begin{lemma}[Interpolation Constraints for $\filtP$ and $\filtM$ \cite{seronFundamentalLimitationsFiltering1997}]\label{lemma:Interpolation_constraints_on_P_and_M}%
	Let  $\rho_i, \, i= 1,\ldots,n_\rho$ be the open RHP poles of $G_{zd}$ and let $\xi_i, \, i= 1,\ldots,n_\xi$ be the open RHP zeros of $G_{yd}$ that are not also zeros of $G_{zd}$. Assume that $F$ is a bounded error estimator. Then 
	\begin{equation}
	\begin{aligned}
	\filtP(\rho_i) &= 0, \quad \filtM(\rho_i) = 1,\\
	\filtP(\xi_i) &= 1, \quad  \filtM(\xi_i) = 0.
	\end{aligned}
	\end{equation}
\end{lemma}

\begin{lemma}[Bode Integral for $\filtP$ \cite{seronFundamentalLimitationsFiltering1997}]\label{lemma:bode_integral_P}%
	Suppose that $\filtP$ is proper and that $F$ is a bounded error estimate.
	Let the open RHP zeros of $\filtP$ be $\varrho_i, \, i= 1,\ldots, n_\varrho$ and let $\varsigma_i, \, i= 1,\ldots, n_\varsigma$ be the open RHP zeros of $G_{zd}$ such that $F(\varsigma_i)G_{yd}(\varsigma_i) \neq 0$. Then if $\filtP(\infty)   \neq 0$
	\begingroup\makeatletter\def\f@size{10}\check@mathfonts
	\def\maketag@@@#1{\hbox{\m@th\normalsize\normalfont#1}}%
	\begin{multline}
	\label{eq:bodeP}
	\int_{0}^{\infty }  \ln \left|\frac{\filtP(\jomega)}{\filtP(\infty)}\right| d\omega = \frac{\pi}{2}\lim\limits_{{s\rightarrow \infty}  }
	\! \! \frac{s\left[\filtP(s)-\filtP(\infty)\right]}{\filtP(\infty) } 	
	\\ 	
	+ \pi \sum_{\hspace{-1cm} i=1 \hspace{-1cm}}^{n_\varrho}\varrho_i - \pi \sum_{\hspace{-1cm} i=1 \hspace{-1cm}}^{n_\varsigma} \varsigma_i\text{\normalsize .}
	\end{multline}
	\endgroup
\end{lemma}

Similar to \cref{lemma:Bode_sensitivity_integral,lemma:Interpolation_constraints_on_S_and_T}, this tells us that the estimation error cannot be made arbitrarily small over all angular frequencies.

\section{Power System Sensor Feedback Limitations}
\label{sec:sensor_feedback_limitations}
The results from \cref{sec:preliminaries} will here be used to identify fundamental sensor feedback limitations for improving rotor angle stability in a two-machine power system.  First, we show that with ideal wide-area measurements, without communication delay, the performance in terms of rotor angle stability is only limited by the available input power. 
Second, we show that using local phase angle measurements, it is impossible to perfectly estimate the excitation of the inter-area mode. Water-bed constraints on the filtering sensitivity dictates that accurate estimation of the inter-area mode has to be compensated by inaccuracy outside a certain angular frequency window. Finally, the main result of the paper is presented. We show that the filtering limitation results in feedback limitation for the closed-loop system using local measurements. Consequently, amplification of certain disturbances are unavoidable.
\subsection{Ideal Feedback Measurement $y=\omega_1-\omega_2$}
\label{sec:fullfeedback}
Inter-area oscillations are an electromechanical phenomenon where groups of machines in one end of the system swing against machines in the other end of the system \cite{kundurPowerSystemStability1994}.
Consider the two-machine system shown in \cref{fig:system_2machine}. 
Here, the inter-area mode is accurately observed from the modal speed \cite{larsenConceptsDesignFACTS1995} which for the two-machine system is given by the relative machine speed 
\begin{equation}
\label{eq:z}
z = \dot \delta_1-\dot \delta_2 =  \omega_1-\omega_2.
\end{equation}
If available, this is the ideal feedback signal \cite{larsenConceptsDesignFACTS1995,bjorkFundamentalPerformanceLimitations2019}.
With input-output mapping given by the transfer function matrix in \eqref{eq:Gdeltaw}
the system to stabilize with feedback control becomes
\begin{equation}
\label{eq:Gzu}
G_{zu} = \frac{X_1^*-X_2^*}{M X_\Sigma^*}\frac{s}{s^2 + \Omega^2}.
\end{equation} 
\begin{prop}
	\label{prop:full_feedback}%
	Suppose that $y=z$, $G_{zu}$ have no open RHP zeros, and that $|G_{zd}|$ roll off at higher angular frequencies. Then for every positive $\epsilon$ there exist a controller $K$ such that
	\begin{equation}
	\label{eq:propTzd}
	|\sensT_{zd}(\jomega)| < \epsilon, \, \forall \omega.
	\end{equation}
\end{prop}
\begin{IEEEproof}%
	First, we note that $u =-K_z z$. Thus, the closed-loop system \eqref{eq:Tzd} reduces to $\sensT_{zd} = \sensS G_{zd}$.
	Since $|\sensT_{zd}| \leq |\sensS||G_{zd}|$, the objective \eqref{eq:propTzd} is fulfilled if
	\begin{equation}
	\label{eq:propTzd_2}
	|\sensS(\jomega)| < \epsilon/|G_{zd}(\jomega)|, \, \forall \omega.
	\end{equation}
	If there are no RHP zeros in $G_{zu}$ then, by \cref{lemma:Interpolation_constraints_on_S_and_T}, $\sensS$ is not constrained at any specific angular frequencies. 
\end{IEEEproof}
\begin{ex}
	\label{ex:perfect_proportional_control}%
	Suppose that we choose proportional control $u = -kz$ then with $G_{zu}$ given by \eqref{eq:Gzu}
	\begin{equation}
	\sensS = \frac{s^2 + \Omega^2}{s^2 + sk\frac{X_1^*-X_2^*}{M X_\Sigma^*} + \Omega^2}.
	\end{equation}
	According to \cref{lemma:Bode_sensitivity_integral} we then have
	\begin{multline}
	\label{eq:bodeInt_S_feedback_with_z}
	\int_{0}^{\infty }  \ln \left|{\sensS(\jomega)} \right| d\omega = \frac{\pi}{2}\lim\limits_{s\rightarrow \infty} s\left[\sensS(s) -1\right] \\  = -k\frac{\pi}{2}\frac{X_1^*-X_2^*}{M X_\Sigma^*}
	\end{multline}
	which, for  $k \left( X_1^*-X_2^* \right)>0$, is always negative.
\end{ex}

This implies that the excitation of the inter-area mode, by load disturbances, can be made arbitrarily small for all angular frequencies. In terms of rotor angle stability, both POD and transient stability are then only limited by the available input power and the achievable actuator bandwidth.

\subsection{Filtering Limitations --- Local Measurement $y=\theta$}
\label{sec:local_frequency}
Typically the industry is restrained from using external communication for crucial system functions such as POD. Thus, using relative machine speed for feedback is normally not an option. The controller instead needs to rely on local measurements.
Here we will show the limitations of estimating the relative machine speed \eqref{eq:z} using local  measurements.

Consider the two-machine system shown in \cref{fig:system_2machine} using local phase angle measurement, $y = \theta$. 
With transfer function 
\begin{equation}
\label{eq:Gzd_Gyd}
\setlength\arraycolsep{2pt}
\begin{bmatrix}
G_{zd_1} & G_{zd_2}\\G_{yd_1} & G_{yd_2}
\end{bmatrix} \! =  
G_0 \! \begin{bmatrix}
s^3  &-s^3
\\
\frac{X_2^*}{X_\Sigma^*}\!\left( s^2 \! + \! \frac{1}{MX_2^*}\! \right)  & \frac{X_1^*}{X_\Sigma^*}\!\left( s^2 \! + \! \frac{1}{MX_1^*}\! \right) \!
\end{bmatrix}\! 
\end{equation}
given by \cref{eq:Gdeltaw,eq:G0_poles,eq:Q_zeros}.
Assume that $0\leq X_1^* < X_2^*\leq X_\Sigma^*$, i.e., machine 1 is closest to the measurement bus. Then the corresponding complex conjugated zero pairs $q_1 = \pm \jimag\sqrt{1/MX_2^*}$ and $q_2 = \pm \jimag\sqrt{1/MX_1^*}$ fulfill
\begin{equation}
	\label{eq:lower_lim_zero}
\Omega/\sqrt{2} \leq |q_1| < |q_2| \leq \infty.
\end{equation}

Partition the general control problem \eqref{eq:Tzd} into a control and an estimation problem as shown in \cref{fig:sensor_feedback}. 
\begin{prop}
	\label{prop:filtering_error}%
	Suppose that $y=\theta$ and let $\filtP_1$ and $\filtP_2$ be the filtering sensitivity functions associated with $d_1$ and $d_2$ respectively. Then an estimator $\hat z = Fy$, such that
	\begin{equation}
	\label{eq:prop_filt_err}
	|\filtP_\idx(\jomega)|	
	<1, \quad \idx = 1,2
	\end{equation}
	is only possible in a frequency interval 
	$\begin{pmatrix} \munderbar \omega, \bar \omega \end{pmatrix} 
	\subset
	\begin{pmatrix}
	|q_1|,|q_2|
	\end{pmatrix}$.
\end{prop}
\begin{IEEEproof}%
	A minimum requirement for $|\filtP_\idx|<1$ is that the estimate $\hat z$ has the same sign as $z$, i.e., that $\filtM_\idx>0$.
	From \eqref{eq:Gzd_Gyd}, the complementary filtering sensitivities to consider becomes
	\begin{equation}
		\label{eq:prop_filtsens}
		\begin{aligned}
			\filtM_1 &= F G_{yd_1}G_{zd_1}^\inv = F\frac{X^*_2}{X_\Sigma^*}\frac{s^2 + |q_1|^2}{s^3},\\
			\filtM_2 &=  F G_{yd_2}G_{zd_2}^\inv = -F\frac{X^*_1}{X_\Sigma^*}\frac{s^2 + |q_2|^2}{s^3}.
		\end{aligned}
	\end{equation}
	Clearly, $\filtM_1(\jimag \omega), \, \filtM_2(\jimag \omega)>0$ are then only possible if $\sign \left(|q_1|^2 - \omega^2\right) = -\sign \left( |q_2|^2 - \omega^2\right) = -1.$
\end{IEEEproof}
\begin{prop}\label{prop:error_amplification}%
	Suppose that $y=\theta$ and that an estimator $\hat z = Fy$ achieves
	\begin{equation}
	\int_{\munderbar \omega}^{\bar \omega} \ln |\filtP_\idx(\jomega)| d\omega <0, \quad \idx =1,2
	\end{equation}
	in a frequency interval 
	$\begin{pmatrix} \munderbar \omega, \bar \omega \end{pmatrix} 
	\subset
	\begin{pmatrix}
	|q_1|,|q_2|
	\end{pmatrix}$. Then amplification of the disturbance on the relative estimation error,
	\begin{equation}
	\label{eq:prop_dist_amp}
	\sup_\omega\, \max \left(|\filtP_1(\jomega)|,|\filtP_2(\jomega)|\right) >1,
	\end{equation}
	is unavoidable in the frequency intervals $\omega< \munderbar \omega$ and $\omega> \bar \omega$.
\end{prop}

\begin{IEEEproof}%
	Since $G_{yd_1}G_{zd_1}^\inv$ and $G_{yd_2}G_{zd_2}^\inv$ in \eqref{eq:prop_filtsens} are strictly proper, $\filtP_\idx(\infty) = 1$. According to \cref{lemma:bode_integral_P}  we then have
	\begin{equation}
	\int_{0}^{\infty}  \ln \left|\filtP_\idx(\jomega)\right| d\omega =
	\frac{\pi}{2}   \lim\limits_{{s\rightarrow \infty}}
	\! {s\left[\filtP_\idx(s)-1\right]} + 
	C_\idx
	\end{equation}
	where 
	$
	C_\idx
	$
	is a non-negative constant resulting from any RHP zeros in $\filtP_\idx$. From \cref{defi:Filtering_Sensitivity}, $\filtP_\idx -1 = -\filtM_\idx$. With the complementary filtering sensitivities given by \eqref{eq:prop_filtsens}, then
	\begin{equation}
	\begin{aligned}
	\int_{0}^{\infty}  \ln \left|\filtP_1(\jomega)\right| d\omega 
	&=-\frac{\pi}{2}  
	F(\infty) \frac{X_2^*}{X_\Sigma^*} +  C_1,
	\\
	\int_{0}^{\infty}  \ln \left|\filtP_2(\jomega)\right| d\omega 
	&=\frac{\pi}{2}  
	F(\infty) \frac{X_1^*}{X_\Sigma^*} +  C_2.
	\end{aligned}
	\end{equation}	
	If the filter is strictly proper, then any region with $|\filtP_\idx(\jomega)|<1$ need to be compensated with an (at least) equally large region with $|\filtP_\idx(\jomega)|>1$. If the filter is proper, this holds true  with respect to at least one of the disturbances $d_1$ or $d_2$.
\end{IEEEproof}

\begin{ex}
	\label{ex:filteringLimitations}%
	Consider the two-machine power system shown in \cref{fig:system_2machine} with linear dynamics \eqref{eq:Gzd_Gyd} derived in \cref{sec:Model}. Let the line reactance $X_\Sigma^* = \SI{1}{\perunit}$ and scale the machine inertia $M$ so that the inter-area modal frequency $\Omega = \sqrt{2/MX_\Sigma^*} = \SI{1}{\radian\per\second}$. In addition, add a \SI{0.05}{\perunit\per(\radian\per\second)} damping constant at each machine so that the inter-area mode has small but positive damping.
	\cref{fig:example1} shows the Bode diagram of $G_{yd_\idx}G_{zd_\idx}^\inv$, $\idx=1,2$, with $X_1^* = \SI{0.1}{\perunit}$ and $X_2^* = \SI{0.9}{\perunit}$, i.e., the control bus is located closer to machine 1. 
	
	Consider the case where the mode is estimated using a simple derivative filter.
	Following the numbers listed in \cref{fig:example1}:
	\begin{enumerate}
		\Myitem The derivative filter adds \SI{90}{\degree} phase to $G_{yd_\idx}G_{zd_\idx}^\inv$ and thus $\filtM_1(\jomega)>0$ for $\omega>|q_2|$. However, disturbances $d_2$ coming from the other end of the system will result in an initial estimate with a \SI{180}{\degree} phase shift, i.e., $\filtM_2(\jomega)<0$. 
		\Myitem At $\omega < |q_1|$ we will have the opposite problem.
	\end{enumerate}
	The mechanical analogy  in \cref{sec:mechanical_analogue} thus appears as a sensor feedback limitation both around the high frequency zero $q_2$ and the low frequency zero $q_1$.
	\begin{enumerate}
		\setcounter{enumi}{2}
		\Myitem In a frequency interval 
		$\begin{pmatrix} \munderbar \omega, \bar \omega \end{pmatrix} 
		\subset
		\begin{pmatrix}
		|q_1|,|q_2|
		\end{pmatrix}$ both transfer functions have the same phase. Thus $\filtM_1,\filtM_2>0$  can be guaranteed no matter the origin of the disturbance.
	\end{enumerate}	
	\begin{figure}[t!]
		\centering
		\includegraphics[scale=1]{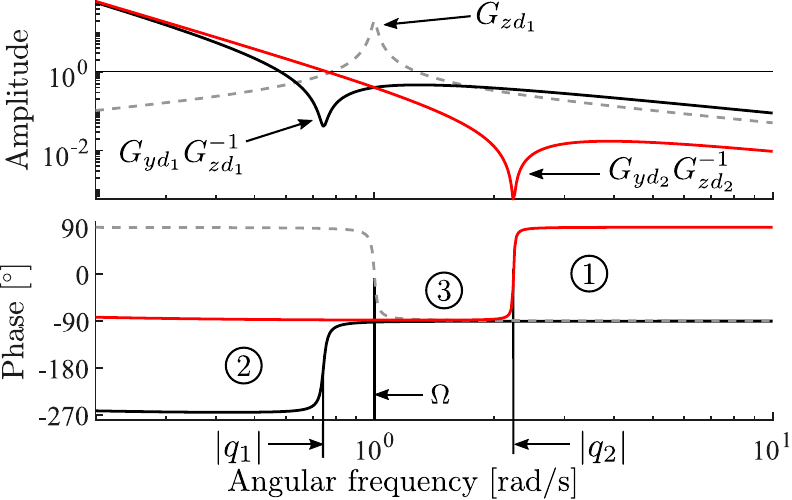}
		\caption{Bode diagram visualizing the filtering limitation when using local phase angle measurement, $y=\theta$, in \cref{ex:filteringLimitations}.}
		\label{fig:example1}
	\end{figure}
\end{ex}

\begin{rem}[Observability Analogy]
	\label{remark:obs}%
	If the control bus are at the electrical midpoint, then the relative frequency $z$ is unobservable using local frequency measurement \cite{smedUtilizingHVDCDamp1993}. At the electrical midpoint, we also have $|q_1| = |q_2| = \Omega$ and thus the frequency interval in which the sign of $z$ can be accurately estimated shrinks to zero. 
	If  the control bus are  adjacent to a machine, then we have best achievable observability of $z$ \cite{smedUtilizingHVDCDamp1993}. Similarly we have the maximum interval, $\begin{pmatrix} \munderbar \omega, \bar \omega \end{pmatrix} 
	\subset
	\begin{pmatrix}
		\Omega/\sqrt{2},\infty
	\end{pmatrix}$. Note that there is still a lower limit, so the considered filtering limitations applies. Ultimately, however, consequences for closed-loop performance are more relevant when the excitation of the mode is large. As seen in \cref{fig:example1}, for $\omega\leq \Omega,$ $|G_{zd}(\jomega)|$ decreases as $\omega\rightarrow 0$. Filtering limitations therefore becomes less significant when measuring closer to a machine.
\end{rem}

\begin{ex}\label{ex:estimation}%
	Consider again the two-machine system introduced in \cref{ex:filteringLimitations}. An estimate of the mode is here obtained by the filter $F_0$, tuned to minimize the $\mathcal{H}_2$-norm from external inputs $d$ and $n$ to the weighted output $e$ as shown in \cref{fig:filterdesign}.
	
	{Note that the $\mathcal{H}_2$ design does not allow for pure integrators in the plant. The pole of the output integral weight is therefore shifted slightly into the LHP. Additionally, the integrator in $G_{yd}$ is canceled in advance by a wash-out filter.}
	
	In the tuning process, external inputs are modeled as white noise with amplitudes $|d_1|, |d_2| = \SI{0.2}{\perunit}$ and $|n|=\SI{0.05}{\radian}$. With the output weight fixed, the ratio $|d|/|n|$ determines the filter performance. 		
	As seen in \cref{fig:stepestimate} the initial sign of $z$ cannot accurately be estimated for both disturbances. By choosing $|d_1|\neq |d_2|$ in the tuning process, we can choose which disturbance to be favored by the filter. The corresponding filtering sensitivities $\filtP_1$ and $\filtP_2$ are shown in \cref{fig:filterd}.
	\begin{figure}[t]
		\centering
		\includegraphics[scale=1.0]{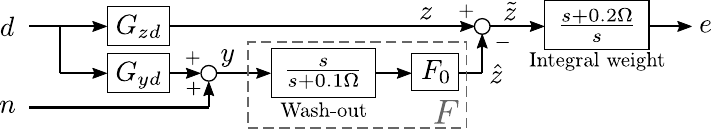}
		\caption{Block diagram for filter design in \cref{ex:estimation}. The integral weight on the output reduces the steady state estimation error.} 
		\label{fig:filterdesign}
	\end{figure}	
	
	\begin{figure}[t!]
		\centering
		\includegraphics[width=\linewidth]{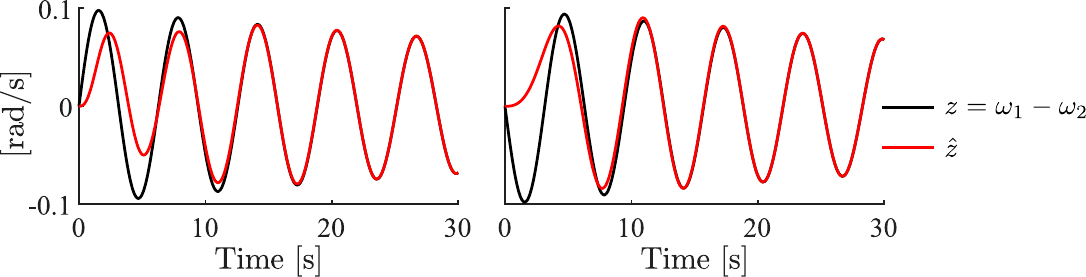}
		\caption{Relative machine speed $z$ and estimate $\hat z$ using local phase angle measurement following a \SI{0.2}{\perunit} \mbox{load step $d_1$ (left) and $d_2$ (right) in \cref{ex:estimation}.}}
		\label{fig:stepestimate}
	\end{figure}
\end{ex}

\subsection{Feedback Limitations --- Local Measurement $y=\theta$}
If there are no RHP zeros or time-delays in $G_{yu}$, then the sensitivity can be made arbitrarily small, e.g., using an inverse based controller. However, making $\sensS$ small is not necessarily the same as making the closed-loop system \eqref{eq:Tzd} small if 
$
y  \neq z.
$
Here it will be shown that the
filtering limitation in Propositions~\ref{prop:filtering_error}~and~\ref{prop:error_amplification} result in closed-loop performance limitations in terms of achievable disturbance attenuation.

The aim of a feedback controller is to reduce the amplitude of the closed-loop system \eqref{eq:Tzd} compared to the open-loop system, i.e., to make
$
|\sensT_{zd}|<|G_{zd}|.
$
Multiplying  with $G_{zd}^\inv$ this can be expressed using the disturbance response ratio 
\begin{equation}
\label{eq:Hzd}
\left|\Rzd\right| = \left|\identity - G_{zu} K (\identity + G_{yu}K)^\inv G_{yd}G_{zd}^\inv\right|  < 1.
\end{equation}

\begin{prop}
	\label{prop:disturbance_attunation}%
	Suppose $y=\theta$ and let $\Rzdone$ and $\Rzdtwo$ \mbox{be the}  disturbance response ratios associated with $d_1$ and $d_2$ respectively. Then a sensor feedback controller $u = -Ky$, such that
	\begin{equation}
	\label{eq:prop_dist_attenuation}
	|\Rzdk(\jomega)|<1, \quad \idx = 1,2
	\end{equation}
	is only possible in a frequency interval 
	$\begin{pmatrix} \munderbar \omega, \bar \omega \end{pmatrix} 
	\subset
	\begin{pmatrix}
	|q_1|,|q_2|
	\end{pmatrix}$.
\end{prop}
\begin{IEEEproof}\label{proof:prop:disturbance_attunation}%
	In \cref{fig:sensor_feedback}, decoupling of the control input $u$ from the estimation error is achieved by selecting $F_u$ such that
	\begin{equation}
	\label{eq:Gzu=Ghat}
	G_{zu} = F(G_{yu} + F_u).
	\end{equation}
	Substituting $G_{yu}$ with $G_{yu}+ F_u$ and $K$ with $K_z F$ then \cref{eq:Hzd} can be written as
	\begin{equation}
	\label{eq:Hzd_2}
	\left|{\Rzdk}\right| = |\identity - 
		{  (\identity +  G_{zu} K_z )^\inv G_{zu}K_z} 
	\filtM_\idx | < 1.
	\end{equation}
	The proof then follows from \cref{prop:filtering_error}.
\end{IEEEproof}
\begin{prop}\label{prop:disturbance_amplification}%
	Suppose that $y=\theta$ and that a sensor feedback controller $u = -Ky$ achieves disturbance attenuation with
	\begin{equation}
	\int_{\munderbar \omega}^{\bar \omega} \ln |\Rzdk(\jomega)| d\omega <0, \quad \idx =1,2
	\end{equation}
	in a frequency interval 
	$\begin{pmatrix} \munderbar \omega, \bar \omega \end{pmatrix} 
	\subset
	\begin{pmatrix}
	|q_1|,|q_2|
	\end{pmatrix}$. Then disturbance amplification
	\begin{equation}
	\label{eq:prop_dist_ampR}
	\sup_\omega\, \max \left(|\Rzdone(\jomega)|,|\Rzdtwo(\jomega)|\right) >1,
	\end{equation}
	is unavoidable in the frequency intervals $\omega<\munderbar \omega$ and $\omega>\bar \omega$.
\end{prop}
\begin{IEEEproof}%
	Bode integral constraints similar to \cref{lemma:bode_integral_P} can be derived also for $\Rzd$.
	Suppose that the closed-loop two-machine system is stable and that the performance variable $z=\omega_1-\omega_2$.
	Since both $G_{zu}$ and $\filtM_\idx$ are strictly proper, $\Rzdk-1$ have a relative degree $\geq 2$.\footnote{ If we instead choose the performance variable as $z = \delta_1-\delta_2$, then $G_{zu}$ have a relative degree of 2 in itself.} 
	Thus,
	\begin{equation}
	\label{eq:waterbed_R}
	\int_{0}^{\infty }  \ln \left|{\Rzdk( \jomega)}\right| d\omega
	= 
	{\pi \sum_{{{r}=1}}^{n_\gamma}\gamma_{r}} \geq 0
	\end{equation}	
	where
	$\gamma_{r}$ are RHP zeros of $\Rzdk$ \cite{freudenbergFundamentalDesignLimitations2003}. 
	It then follows that disturbance amplification \cref{eq:prop_dist_ampR} is unavoidable in the frequency intervals $\omega<\munderbar \omega$ and $\omega>\bar \omega$ due to the water-bed effect.
\end{IEEEproof}
\begin{rem}%
	RHP zeros of $\filtP_\idx$ are not necessarily shared with $\Rzdk$. But
	if $K_z\rightarrow \infty$ in \eqref{eq:Hzd_2} then $\Rzdk\rightarrow \filtP_\idx$.
\end{rem}
\begin{ex}
	\label{ex:feedback_lim}%
	Consider again the two-machine system with the filter $F$ designed in \cref{ex:estimation}. 
	Since the filter gives an estimate of the mode, damping of the mode can be improved by closing the loop with a controller $K_z = \SI{0.5}{\perunit/(\radian\per\second)}$ as shown in 
	\cref{fig:sensor_feedback}.
	In accordance with
	Propositions \ref{prop:disturbance_attunation}~and~\ref{prop:disturbance_amplification} this result in an unavoidable disturbance amplification outside the frequency interval $\begin{pmatrix} \munderbar \omega, \bar \omega \end{pmatrix} 
	\subset
	\begin{pmatrix}
	|q_1|,|q_2|
	\end{pmatrix}$
	for either $\Rzdone$ or $\Rzdtwo$ as seen in \cref{fig:filterd}.

	\begin{figure}[t!]
		\centering
        \includegraphics[width=1\linewidth]{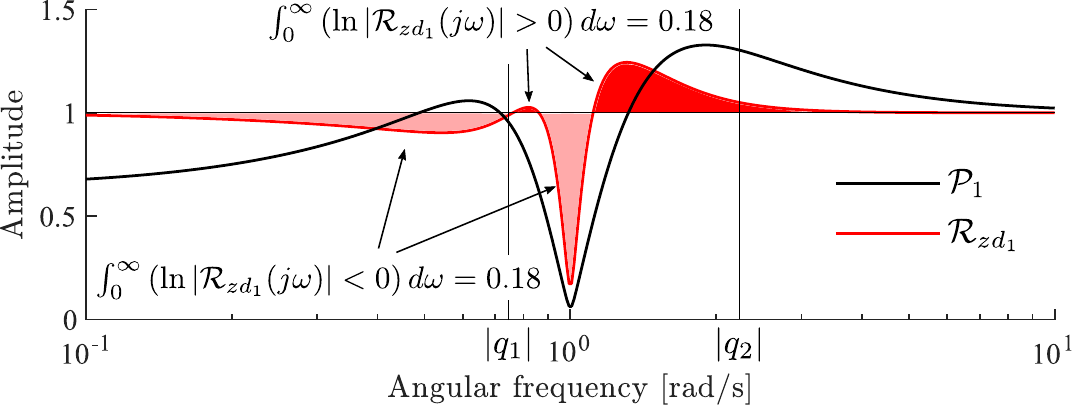}
        \\
        \vspace{2mm}
		\includegraphics[width=1\linewidth]{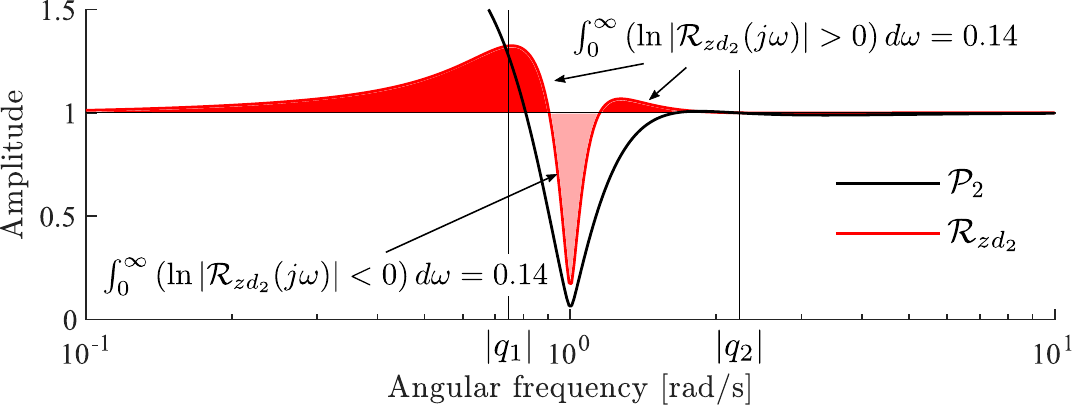}
		\caption{Filtering sensitivity and disturbance response ratio with respect to $d_1$ (top) and $d_2$ (bottom) in Examples~\ref{ex:estimation}~and~\ref{ex:feedback_lim}. Due to the water-bed effect, disturbance amplification is unavoidable around the zero frequencies $|q_1|$~and~$|q_2|$.}
		\label{fig:filterd}
	\end{figure}
\end{ex}

\section{Nonlinear Simulation Study}
\label{sec:Simulink}
\begin{figure}[t]
	\centering
	\includegraphics[width= 1\linewidth]{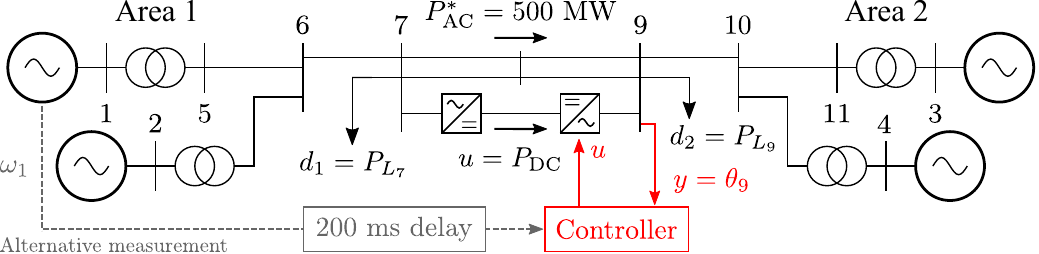}
	\caption{Four-machine two-area system \cite{kundurPowerSystemStability1994,kamwaPerformanceThreePSS}. Modifications: embedded \HVDClink, inertia reduced to 75\%, inter-area ac power flow increased to \SI{500}{\mega \watt}, and PSSs tuned down for a marginally damped inter-area mode.}
	\label{fig:kundur_system}
\end{figure}

In this section, we study the filtering limitations in a Simulink implementation \cite{kamwaPerformanceThreePSS} of the Kundur four-machine two-area test system \cite{kundurPowerSystemStability1994} shown in \cref{fig:kundur_system}. For illustrative purposes, the system has been modified by reducing the POD. 
We assume the controllable device is an embedded \HVDClink~where the dc power can be controlled in a linear region with sufficiently high bandwidth. Without further loss of generality, the dc dynamics are then neglected, as motivated in \cite{harneforsImpactInterareaModes2017}.
We begin this section by showing the filtering limitations
associated with local phase angle or frequency measurement. Then we show the consequences this have on rotor angle stability. Finally, it is shown how alternative measurement signals can be useful to limit rotor angle stability issues associated with phase angle measurements in the considered four-machine test system.

To facilitate the analysis and controller design, we need a linear representation of the system. Using the Simulink linear analysis toolbox, we obtain the 90\textsuperscript{th} order state-space realization
\begin{equation}
	\label{eq:Linearized}
	\begin{aligned}
		\dot x &= Ax + B_d d + B_u u \\
		y &= C_yx + D_{yd}d + D_{yu}u \\
		z &= C_z x
	\end{aligned}
\end{equation}
where $x$ is the state vector representing rotor speeds, rotor angles, generator voltages, controller states, etc.; $d$ is a vector of external inputs, such as the active power load at buses 7 and 9; $u$ is the controllable dc power in the \HVDClink; $y$ is some measured output such as the phase angle $\theta_9$; and $z$ is a performance variable chosen to represent the inter-area mode. 

\subsection{Filtering Limitations --- Local Measurement $y=\theta_9$}
\label{sec:sim_filtering_lim}
Let us illustrate how to extend the analysis in \cref{sec:local_frequency} to study the filtering limitations associated with local phase angle measurement in the four-machine system.
To do this, it is convenient to represent \eqref{eq:Linearized} on its modal form.

Let $\lambda_i, \ i = 1,\ldots,90$ be the eigenvalues of $A$ so that $\det(A-\lambda_i I) = 0$. Let $V = \begin{bmatrix}
	v_1, \ldots, v_{90}
\end{bmatrix}\in\complex^{90\times 90}$ and $U = \begin{bmatrix}
u_1, \ldots, u_{90}
\end{bmatrix}\in\complex^{90\times 90}$ be matrices of left and right eigenvectors so that
$v_i^\Tconj A = v_i^\Tconj \lambda_i$, $ Au_i = \lambda_i u_i$, and $V^\Tconj = U^\inv$, where $v_i^\Tconj$ is the complex conjugate transpose of $v_i$. 

Transforming the state-space coordinates $x$ into the modal coordinates $\zeta = V^\Tconj x$ we can rewrite \eqref{eq:Linearized} as
\begin{equation}
	\label{eq:Linearized_diag}
	\begin{aligned}		
	\dot \zeta &= V^\Tconj A U \zeta + V^\Tconj B_d d + V^\Tconj B_u u 
	\\
	y &= C_y U \zeta + D_{yd} d + D_{yu} u 
	\\
	z &= C_z U \zeta
	\end{aligned}
\end{equation}
where $V^\Tconj A U  = \diag(\lambda_1,\ldots, \lambda_{90})\in\complex^{90\times90}$.

Oscillatory modes, $\lambda_i$, are characterized by two complex conjugated modal states. For instance let the oscillatory inter-area mode be represented by the pair $\zeta_1$ and $\zeta_2 = \bar\zeta_1$.
Ideally, to stabilize the mode, we would like a good estimate of either $\zeta_1$ or $ \zeta_2$.
The problem is that most available numerical methods do not allow for a complex valued state-space representation. This can be amended by instead representing \cref{eq:Linearized_diag} on real Jordan form \cite{hornTopicsMatrixAnalysis1991}.
Let $\bm V = \begin{bmatrix}
	\bm v_1, \ldots, \bm v_{90}
\end{bmatrix}\in\real^{90\times 90}$ be a transformation matrix with 
\begin{itemize}
	\item \label{item_proj}
	$\begin{bmatrix}
		\bm v_i, \bm v_{i+1}
	\end{bmatrix} = \begin{bmatrix}\Real v_i, \Imag v_i\end{bmatrix}$ if  $\lambda_{i} = \bar \lambda_{i+1}$, and 
	\item $\bm v_i = v_i$ if $\lambda_i$ is real.
\end{itemize} 

With the coordinate transform $\bm \zeta = \bm V^\T x$, we rewrite the system \eqref{eq:Linearized_diag} on real Jordan form
\begin{equation}
	\label{eq:Linearized_block_diag}
	\begin{aligned}		
		\dot {\bm\zeta} &= \bm V^\T A \bm U \bm\zeta + \bm V^\T B_d d + \bm V^\T B_u u 
		\\
		y &= C_y \bm U \bm \zeta + D_{yd} d + D_{yu} u 
		\\
		z &= C_z \bm U \bm \zeta
	\end{aligned}
\end{equation}
where $\bm U^\inv =\bm V^\T$ and $\bm V^\T A \bm U \in \real^{90\times90}$ is block-diagonal.

The state-space realization \eqref{eq:Linearized_block_diag} can be used to study the filtering limitations associated with local phase angle measurements. But first, to facilitate the subsequent controller design, let us rotate the eigenvectors so that $\bm \zeta_1$ and $\bm \zeta_2$ are suitable to represent the damping and synchronizing torque of the inter-area mode. 

\begin{figure}
	\centering
	\includegraphics[scale=0.4]{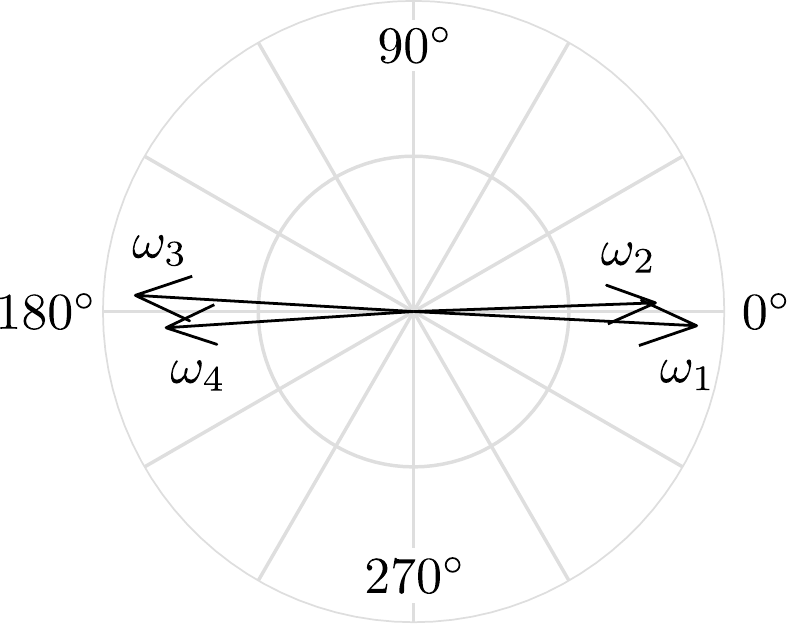}
	\caption{Mode shape of the inter-area mode in the four-machine system.}
	\label{fig:compas}
\end{figure}

For the undamped symmetrical two-machine system \eqref{eq:Gdeltaw}, suitable states to represent the damping and synchronizing torque would be
$\bm \zeta_1 = \omega_1-\omega_2$ and $\bm \zeta_2 = \delta_1-\delta_2$ respectively. 
Suitable states to represent the damping and synchronizing torque in the detailed four-machine model \cref{eq:Linearized} are not as obvious. However, if we assume that the inter-area oscillations are still dominated by the mechanical dynamics of the synchronous machines we can rotate the corresponding eigenvectors, e.g., $v_1$ and $v_2 = \bar v_1$, so that the elements corresponding to machine speeds are aligned with the real axis, as shown in \cref{fig:compas}. This makes
\begin{equation}
	\label{eq:damping_and_sync_zeta}	
	\bm \zeta_1 = \Real  v_1^\Tconj  x = \bm v_1^\T  x \quad \text{and} \quad \bm \zeta_2 = \Imag  v_1^\Tconj x = \bm v_2^\T  x
\end{equation}
state variables suitable for analyzing damping and synchronizing torque in the four-machine system even though governors, voltage dynamics, higher-order generator dynamics, etc., have been introduced \cite{bjorkFundamentalControlPerformance2021}. 
To target damping of the inter-area mode, let the performance variable
\begin{equation}
	\label{eq:performance_variable}
	z = \bm \zeta_1 =  C_z x = \bm v_1^\T  x
\end{equation}
where $v_1$ is the left eigenvector associated with the inter-area mode, aligned with the real axis as shown in \cref{fig:compas}.

\begin{figure}[t!]
		\centering
		\includegraphics[width=0.95\linewidth]{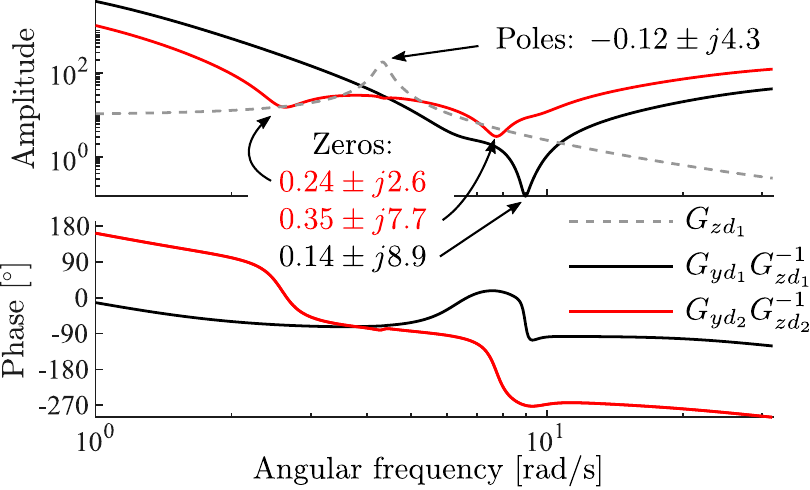}
		\caption{Bode diagram visualizing the filtering limitation with local phase angle measurement, $y = \theta_9$, and load disturbances $d_1$ and $d_2$ at buses 7 and 9 respectively, as shown in \cref{fig:kundur_system}.}
		\label{fig:kundur_filter}
	\end{figure}

	Assume now that we use local measurement $y=\theta_9$ to improve POD in the four-machine system shown in \cref{fig:kundur_system}. 		
	The sensor feedback limitations considering active power disturbances $d_1$ and $d_2$ at buses 7 and 9, respectively, are shown in \cref{fig:kundur_filter}. The Bode diagram bares close resemblance to the simplified model in \cref{fig:example1}. 
	The main difference is that $G_{yd_2}G_{zd_2}^\inv$ shows a prominent zero pair also at higher angular frequencies. This is because the disturbance $d_2$ occur at the measurement bus. As shown in \eqref{eq:Gdeltaw}, the transfer function $G_{yd_2}$ therefore has a low frequency and a high frequency zero pair. \cref{prop:error_amplification} implies that it will be impossible to accurately determine the source of the load disturbance.
	According to \cref{prop:disturbance_amplification}, POD using local frequency measurements will therefore unavoidably amplify disturbances around the low  frequency and high frequency zero pairs. In the following subsection, we show the consequence of this for transient rotor angle stability.
	
	It is also worth noting that the electromechanical zero pairs are in the RHP. This is often the case in power systems due to the destabilizing effect of generator excitation controllers~\cite{bjorkControlLimitationsDue2020}. Since we have RHP zeros, interpolation constraints on the sensitivity further limits the achievable closed-loop performance, according to \cref{lemma:Interpolation_constraints_on_S_and_T}. Ultimately, this will have consequences for robustness. It is therefore advisable not demand too much from the feedback controller \cite{skogestadMultivariableFeedbackControl2007}. 
	
\subsection{Feedback Limitations --- Local Measurement $y=\theta_9$}
\label{sec:simulation_feedbacl_lim}
To show the consequences of the filtering limitations on rotor angle stability, we first need to design a feedback controller. In this section, we will compare two types of controllers. A conventional phase compensating controller, typically used for PSS implementations, and a $\mathcal{H}_2$ optimal controller tuned to attenuate the performance variable \eqref{eq:performance_variable}.

\subsubsection{PSS-Style Controller}
\label{sec:pss-style}

\begin{figure}[t]
	\centering
	\includegraphics[width=1\linewidth]{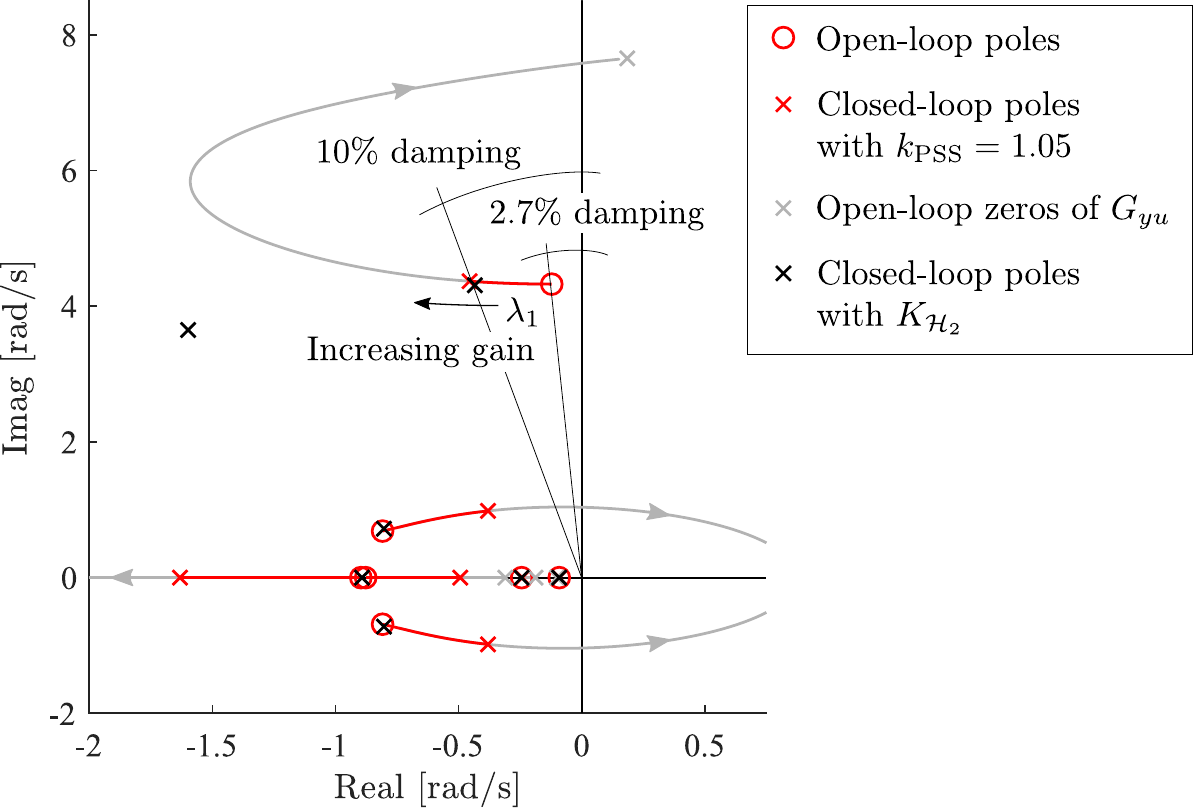}
	\caption{Root locus used for selecting controller gain $k_\textrm{PSS}$.}
	\label{fig:rootlocus}
\end{figure}

Let  $P_\mathrm{DC}=-K_\textrm{PSS} \theta_9$
\begin{equation}
	\label{eq:K_Kpss}
	K_\textrm{PSS} = s \, k_\textrm{PSS}
	\underbrace{\frac{sT_1+1}{sT_2+1}}_{\mathclap{\textrm{Phase compensation} \quad}} 	
	{\underbrace{\left(
	\frac{5 \Omega_1}{s + 5 \Omega_1}
	\right)}_{{\textrm{Low-pass}}} }^2 
	\underbrace{\frac{s}{s+0.2 \Omega_1}}_{\textrm{Wash-out}} 
\end{equation}
where $\Omega_1 = |\lambda_1| \approx \SI{4.4}{\radian\per\second}$ is the frequency of the poorly damped inter-area mode.
The eigenvalue sensitivity to small changes in $K_\textrm{PSS}$ is given by the residue $R(\lambda_1)= -C_y u_1v_1^\Tconj B_u$ \cite{pagolaSensitivitiesResiduesParticipations1989}.
We tune the phase compensation in \eqref{eq:K_Kpss} so that $\arg R(\lambda_1) K(\lambda_1) = -\pi$. Thus, feedback moves the eigenvalue, $\lambda_1$, associated with the inter-area mode in the negative real direction as shown in \cref{fig:rootlocus}. Choosing $k_\textrm{PSS}=1.05$, the damping of the inter-area mode improves from \SI{2.7}{\percent} to \SI{10}{\percent}. 
The low-pass filters in \eqref{eq:K_Kpss} are selected to avoid amplification of high frequency measurement noise. The wash-out filter is tuned to avoid amplification at low angular frequencies.
The required phase compensation, $-\SI{5}{\degree}$, is achieved with $T_1 = 0.21$ and $T_2 = 0.25$ in \cref{eq:K_Kpss}.  

\begin{figure}
	\centering
	\includegraphics[width=1\linewidth]{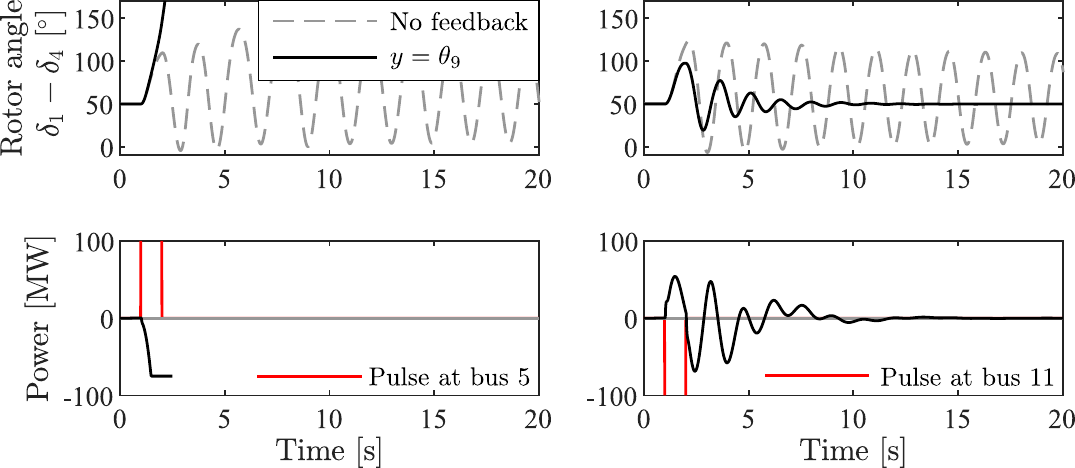}
	\caption{Rotor angle differences and control input following a \SI{350}{\mega\watt} disturbance pulse  with $P_\textrm{DC} = - K_\textrm{PSS}\theta_9$. }
	\label{fig:kpss}
\end{figure}

In \cref{fig:kpss}, the system response to a \SI{1}{\second} long \SI{350}{\mega\watt} disturbance is shown. The worst case disturbances are those that increase the rotor phase angle difference. Therefore we consider a load loss at bus 5 and a generation loss at bus 11
(simulated as active power load pulses). This could for example represent the commutation failure of an exporting or importing \HVDClink. 

The trajectory of $\lambda_1$ in \cref{fig:rootlocus} indicates that the implemented controller improves damping torque. It also marginally improves synchronizing torque. This could lead us to believe that both POD and transient stability have been improved. Indeed, as seen in \cref{fig:kpss}, both the damping and transient response are improved for disturbances occurring close to the measurement bus at bus~11. However, for disturbances occurring in the other end of the system, this is not the case. In accordance with \cref{prop:disturbance_amplification}, an erroneous control input  and thus a disturbance amplification is unavoidable. This increases the initial angle, $\delta_1-\delta_4$, leading to a system separation in the first swing. 

\subsubsection{$\mathcal{H}_2$ Optimal Controller}
\label{sec:H2controller_design}%
\begin{figure}
	\centering
	\includegraphics[scale=1]{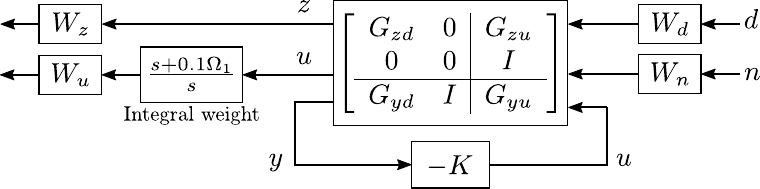}
	\caption{Extended system for optimal control design. }
	\label{fig:controldesign}
\end{figure}

Let $P_\mathrm{DC} = - K_{\mathcal{H}_2} \theta_9$, where $K_{\mathcal{H}_2}$ is the controller that minimizes the $\mathcal{H}_2$-norm from white noise inputs $d$ and $n$ to weighted outputs in the extended system shown in \cref{fig:controldesign}. Let power and phase be given in \si{\mega\watt} and degrees, respectively, and
\begin{itemize}
	\item let load disturbances $|d| = 1$ and measurement noise $ |n| = 1$, and let the input weights $W_d/W_n = 100$;
	\item let $W_u=1$; and
	\item let $|C_z| = 1$ in \eqref{eq:performance_variable} and $W_z = 40\cdot360$.
\end{itemize}
The tuned controller then achieves \SI{10}{\percent} damping of the inter-area mode. 
Closing the feedback loop, the eigenvalues of the inter-area mode are shifted into the LHP whereas the remaining open-loop poles are unaffected, as shown in \cref{fig:rootlocus}. The optimal controller has the same order as the extended system. Here however, it can be reduced to 4\textsuperscript{th} order without significantly changing the dynamical closed-loop performance. The implemented controller is 
\begin{equation}
\label{eq:KH2_controller}
	K_{\mathcal{H}_2} =
	\frac{0.042 s (172-s) (s+12.6) (s+0.59)}{(s^2 + 3.9s + 19.4) (s^2 + 3.4s + 51.1)}.
\end{equation}

\begin{rem}%
	Note that the reduced controller \eqref{eq:KH2_controller} has a RHP zero at $s=172$. However, since $\Omega_1 \ll 172$, this is insignificant for the closed-loop performance.
\end{rem}

\begin{figure}
	\centering
	\includegraphics[width=1\linewidth]{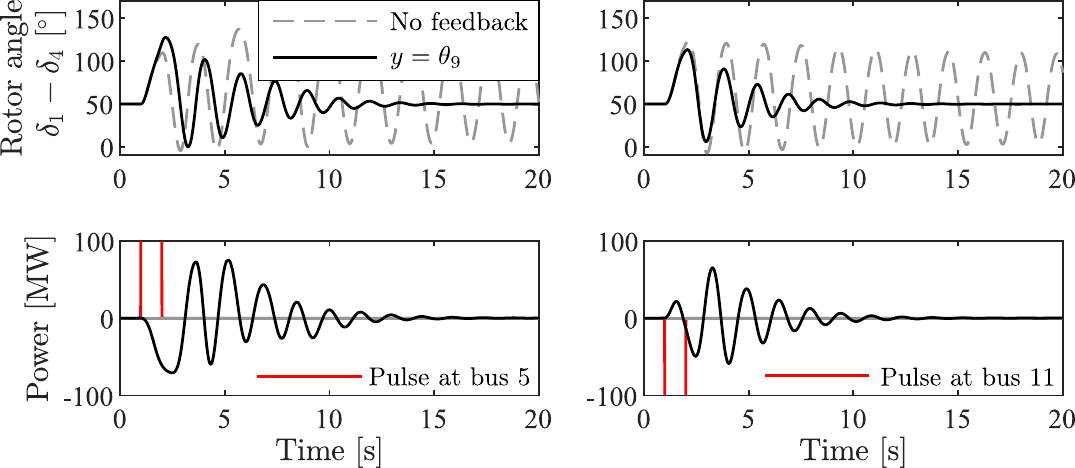}
	\caption{Rotor angle differences and control input following a \SI{350}{\mega\watt} disturbance pulse with $P_\textrm{DC} = - K_{\mathcal{H}_2}\theta_9$. }
	\label{fig:kH2}
\end{figure}

In \cref{fig:kH2}, the system response to a \SI{1}{\second} long \SI{350}{\mega\watt} disturbance is shown. Just as in \cref{fig:kpss} we see good damping as well as improved transient performance when disturbances occur close to the measurement bus. Unlike the PSS-style controller, we see that the $\mathcal{H}_2$ optimal controller survive the first swing, also for disturbances occurring at bus 5. This implies that transient stability issues may be avoided with proper controller tuning. However, also the well tuned  $\mathcal{H}_2$ optimal controller amplifies the first swing as seen in \cref{fig:kH2}. In accordance with \cref{prop:disturbance_amplification}, this is unavoidable. 

For the simulations, the controllable dc power variation has been saturated at $\pm\SI{75}{\mega\watt}$ from the steady state set point. One way to reduce the erroneous excitation of the inter-area mode is to reduce this saturation limit. This would improve the transient stability margin for disturbances occurring in the other end of the system.
It is also worth noting that both the PSS-style controller and the $\mathcal{H}_2$ controller improve POD as well as first swing stability in the case where load disturbances occur in the same area. 
In a more distributed setting where POD is provided by decentralized controllers in each end of the system, then the negative effect on the first swing will be counterbalanced by controllers in the other end of the system.

\begin{rem}[Weight Selection]%
	There are many ways of choosing the input and performance weights. For instance, the input weights can be chosen to reflect the size and dynamics of the expected load disturbances and to account for PMU measurement noise. Similarly, the performance weights can reflect the allowed signal sizes and input usage. This is particularly useful in $\mathcal{H}_\infty$ control design since then performance specifications can be considered together with the small-gain theorem, guaranteeing robust performance by ensuring that the $\mathcal{H}_\infty$-norm of the closed-loop is less than one. 
	A convenient design procedure is often to first design the controller by optimizing the $\mathcal{H}_2$-norm, and then to assess robustness using the $\mathcal{H}_\infty$-norm \cite{skogestadMultivariableFeedbackControl2007}. 
\end{rem}
	
\subsection{Feedback Limitations --- Alternative Measurements}
\label{sec:improve_transient_stab}
In this section we will consider $\mathcal{H}_2$ optimal feedback control with alternative measurements. It will be shown that by using WAMS or by substituting local phase angle measurement with local power flow or voltage measurements, the transient stability issues associated with local phase angle measurements can be avoided. Conclusions from the analysis in \cref{sec:sensor_feedback_limitations} can, to some extent, be extrapolated to power flow and voltage measurements.

\subsubsection{WAMS}
\label{sec:wide-area}
\begin{figure}
	\centering
	\includegraphics[width=1\linewidth]{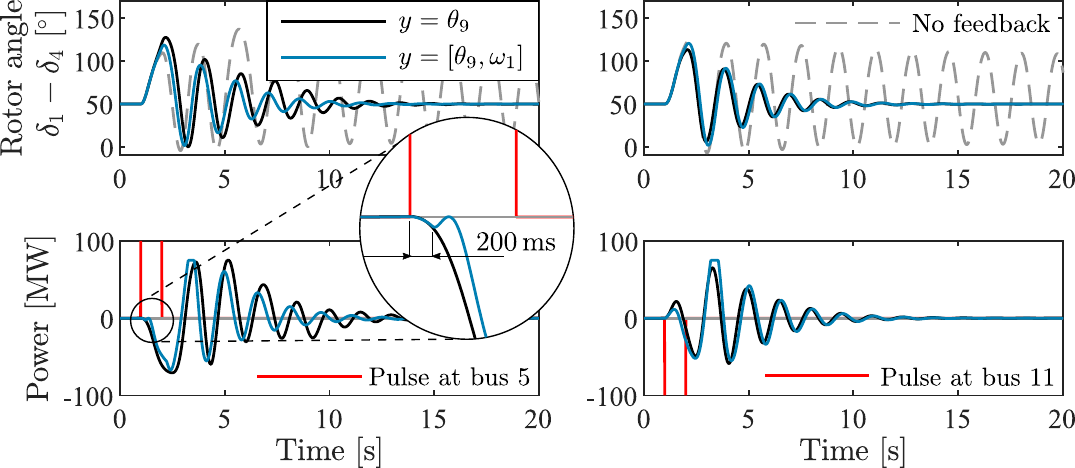}
	\caption{Rotor angle differences and control input following a \SI{350}{\mega\watt} disturbance pulse. Transient rotor angle stability is improved using WAMS. }
	\label{fig:k2w1}
\end{figure}

System awareness can be improved by using WAMS. Complementary measurements should be selected at buses with good observability, and ideally at the other end of the system, to provide as much information as possible. To complement local phase angle measurement, one suitable candidate is therefore the rotor speed  $\omega_1$. Assume a \SI{200}{\milli\second} communication delay. Using the $\mathcal{H}_2$ optimal control design method in \cref{sec:H2controller_design} the controller is tuned to achieve \SI{10}{\percent} damping of the inter-area mode. As shown in, \cref{fig:k2w1}, the use of {WAMS} improves the transient performance by reducing the amplification of the first swing.

\subsubsection{AC Power Flow $y = P_\textrm{AC}$}
Let the feedback controller be an $\mathcal{H}_2$ optimal controller tuned to improve the damping to \SI{10}{\percent} using measurement of ac power flow in one of the lines between buses 8 and 9.

\begin{figure}
	\centering
	\includegraphics[width=0.95\linewidth]{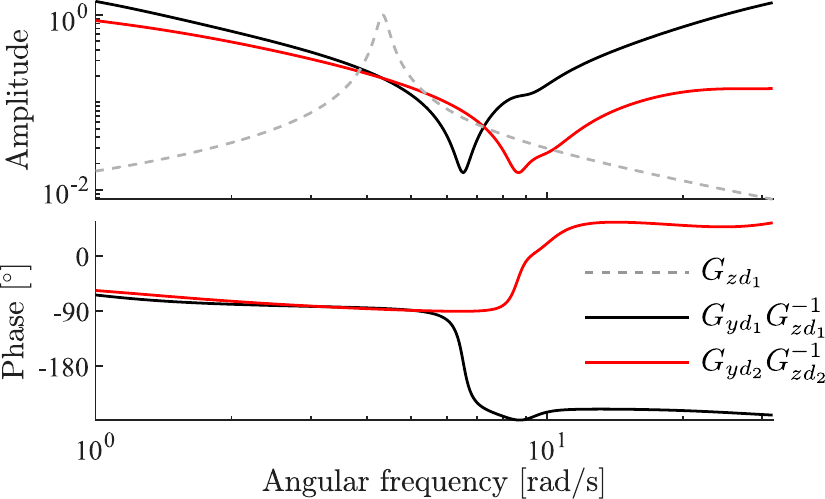}
	\caption{Bode diagram visualizing the complementary filtering sensitivity with local power flow measurement, $y=P_\mathrm{AC}$, and load disturbances $d_1$ and $d_2$ at buses 7 and 9 respectively, as shown in \cref{fig:kundur_system}. Compared to \cref{fig:kundur_filter} we see that measurement of ac power flow does not show the same filtering limitations associated with local phase angle measurement.}
	\label{fig:filter_Pac}
\end{figure}

\begin{figure}[t!]
	\centering
	\includegraphics[width=1\linewidth]{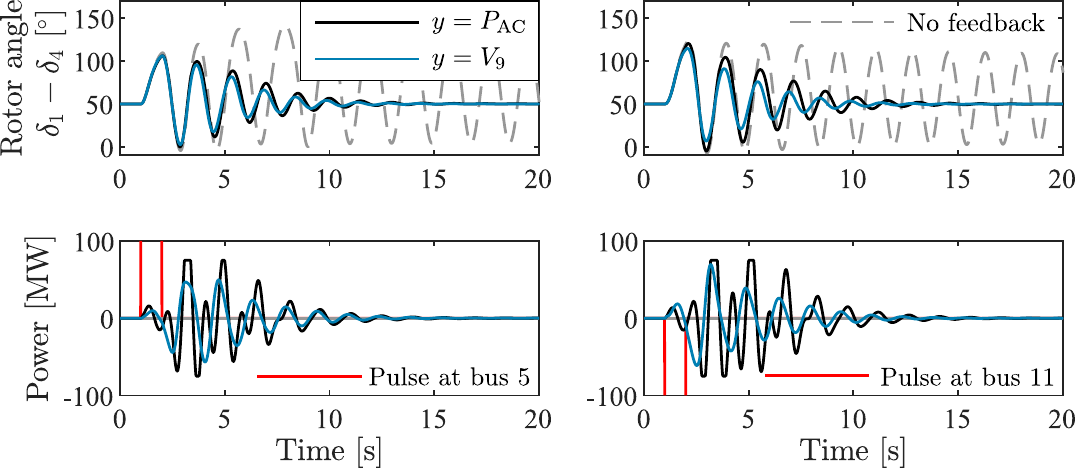}
	\caption{Rotor angle differences and control input following a \SI{350}{\mega\watt} disturbance pulse. For the considered two-area system, feedback control using ac power flow or local voltage measurement does not show the same issues with transient rotor angle stability as feedback from local phase angle measurement. }
	\label{fig:kPacV}
\end{figure}

Since the power flow over the line depends on the angle difference, $\theta_9-\theta_7$, it is useful for determining  the source of the disturbance. 
This can be confirmed by studying the complementary filtering sensitivity in \cref{fig:filter_Pac}. Therefore, as seen in \cref{fig:kPacV}, power flow does not exhibit the same issues with transient stability as local phase angle measurements.
For this reason, ac power flow would here be a better measurement choice compared to phase angle measurements. For a more complex system, however, this may not necessarily be the case as observability will depend on the topology of the particular system. For instance, the initial surge of power will always be in the direction of the load disturbance. If measuring in a shunt, e.g. the electric power of a synchronous machine, then we may experience similar filtering limitations as with phase angle measurements. 
Power flow also shows more sensitivity to process noise. In \cref{fig:kPacV}, although the attenuation of the rotor angle looks smooth, there is quite a bit of fluctuation in the input signal due to the interaction with the local modes.
To extend the analysis, it will be interesting to consider a more detailed network model.

\subsubsection{Bus Voltage $y = V_9$}
Let the feedback controller be an $\mathcal{H}_2$ optimal controller tuned to improve the damping to \SI{10}{\percent} using measurement of local bus voltage $V_9$. As seen in \cref{fig:kPacV}, using local voltage measurement, the feedback controller efficiently attenuates disturbances originating in either end of the system. 

Typically, voltage measurement is best used in combination with reactive power control, e.g., from HVDC converters \cite{smedUtilizingHVDCDamp1993} or FACTS \cite{larsenConceptsDesignFACTS1995,mhaskarPowerOscillationDamping2006}, but it could also be used when controlling active power. Best observability of the inter-area mode is achieved when measuring the voltage close to the electrical midpoint.
As seen in \cref{fig:kPacV}, POD control implemented using voltage measurement is less sensitive to the location of active power load disturbances. 
The reason for this is that the average speed mode is not observable in the voltage. But there could be other dynamical phenomena, for instance, involving reactive power disturbances and short circuits faults that could exhibit similar first swing stability issues. The classical machine model \eqref{swing:eq2} used for the analysis does not capture voltage dynamics. To properly extend the analysis, a more detailed machine modeling is needed.
It is also worth noting that the use of voltage measurement can have other drawbacks, not captured in this study. For instance, when using voltage measurement in combination with reactive power control, the residue angle is sensitive to load dynamics \cite{larsenSVCControlDesign1987}. This can have consequences for the robustness of the control design.

\section{Conclusions}
\label{sec:conclusion}
Sensor feedback limitations for improving rotor angle stability using feedback from local phase angle measurements have been studied. For a linearized two-machine power system model it was shown that, although arbitrarily good damping can be achieved, the sensor feedback limitation dictates that damping improvement must come at the cost of decreased transient performance. Using a detailed power system model, it was shown that this decrease in transient performance may result in transient instability. 

The aim of this work has been to introduce a new analysis method to bridge the gap between small-signal analysis and the study of fundamental performance limitations. In the transition towards more converter-based renewable energy, the importance of control is likely to become even more important than it is today. More understanding of the fundamental limitations will be useful when choosing the control architecture and when designing control algorithms.

It was shown that transient stability can be improved by using wide-area measurements.
In our future work, we will study how communication time-delays affect the potential benefit of complementary wide-area measurements.
It was also shown that transient stability margin was improved for disturbances originating in the same end of the network. Future work will extend this analysis to study how coordinated tuning of distributed controllers can be used to improve POD without communication, while still ensuring sufficient transient stability margins. In the simulation study, it was also shown that transient stability could be preserved by substituting phase angle measurement with power flow or voltage measurements. At least in the considered example.
Future work will extend the analysis to involve voltage dynamics and to consider a more general multi-machine system. This allows for bus voltage measurements and a more accurate representation of the ac power flow. Additionally, filtering limitations associated with reactive power disturbances and short circuit faults may also be considered.